\documentclass[aps,prc,showpacs,reprint,superscriptaddress,letterpaper]{revtex4-1}
\usepackage{graphicx}
\usepackage{bm}
\usepackage{amsfonts}
\usepackage{amsmath}
\usepackage{amssymb}
\usepackage{dcolumn}
\usepackage[autolanguage]{numprint}
\usepackage{color}
\usepackage[table]{xcolor}
\usepackage{psfrag}
\begin{document}

\preprint{V. 0.5}

\title{${\bm \beta}$-delayed $\gamma$ decay of $\bm{^{26}\mathrm{P}}$: Possible evidence of a proton halo}


\author{D. P\'erez-Loureiro}
\email[]{perezlou@nscl.msu.edu}
\affiliation{National Superconducting Cyclotron Laboratory, Michigan State University, East Lansing, Michigan 48824, USA}
\author{C. Wrede}
\email[]{wrede@nscl.msu.edu}
\affiliation{National Superconducting Cyclotron Laboratory, Michigan State University, East Lansing, Michigan 48824, USA}
\affiliation{Department of Physics and Astronomy,  Michigan State University, East Lansing, Michigan 48824, USA}
\author{M.~B.~Bennett}
\affiliation{National Superconducting Cyclotron Laboratory, Michigan State University, East Lansing, Michigan 48824, USA}
\affiliation{Department of Physics and Astronomy, Michigan State University, East Lansing, Michigan 48824, USA}
\author{S.~N.~Liddick}
\affiliation{National Superconducting Cyclotron Laboratory, Michigan State University, East Lansing, Michigan 48824, USA}
\affiliation{Department of Chemistry, Michigan State University, East Lansing, Michigan 48824, USA}
\author{A.~Bowe}
\affiliation{National Superconducting Cyclotron Laboratory, Michigan State University, East Lansing, Michigan 48824, USA}
\affiliation{Department of Physics and Astronomy, Michigan State University, East Lansing, Michigan 48824, USA}
\affiliation{Physics Department, Kalamazoo College, Kalamazoo, Michigan 49006, USA}
\author{B. A. Brown}
\affiliation{Department of Physics and Astronomy,  Michigan State University, East Lansing, Michigan 48824, USA}
\affiliation{National Superconducting Cyclotron Laboratory, Michigan State University, East Lansing, Michigan 48824, USA}
\author{A.~A.~Chen}
\affiliation{Department of Physics and Astronomy, McMaster University, Hamilton, Ontario L8S 4M1, Canada}
\author{K.~A.~Chipps}
\affiliation{Department of Physics, Colorado School of Mines, Golden, Colorado 08401, USA}
\affiliation{Physics Division, Oak Ridge National Laboratory, Oak Ridge, Tennessee 37831, USA}
\affiliation{Department of Physics and Astronomy, University of Tennessee, Knoxville, Tennessee 37996, USA}
\author{N.~Cooper}
\affiliation{Department of Physics and Wright Nuclear Structure Laboratory, Yale University, New Haven, Connecticut 06520, USA}
\author{D.~Irvine}
\affiliation{Department of Physics and Astronomy, McMaster University, Hamilton, Ontario L8S 4M1, Canada}
\author{E.~McNeice}
\affiliation{Department of Physics and Astronomy, McMaster University, Hamilton, Ontario L8S 4M1, Canada}
\author{F.~Montes}
\affiliation{National Superconducting Cyclotron Laboratory, Michigan State University, East Lansing, Michigan 48824, USA}
\affiliation{Joint Institute for Nuclear Astrophysics, Michigan State University, East Lansing, Michigan 48824, USA}
\author{F.~Naqvi}
\affiliation{Department of Physics and Wright Nuclear Structure Laboratory, Yale University, New Haven, Connecticut 06520, USA}
\author{R.~Ortez}
\affiliation{Department of Physics and Astronomy, Michigan State University, East Lansing, Michigan 48824, USA}
\affiliation{National Superconducting Cyclotron Laboratory, Michigan State University, East Lansing, Michigan 48824, USA}
\affiliation{Department of Physics, University of Washington, Seattle, Washington 98195, USA}
\author{S.~D.~Pain}
\affiliation{Physics Division, Oak Ridge National Laboratory, Oak Ridge, Tennessee 37831, USA}
\author{J.~Pereira}
\affiliation{National Superconducting Cyclotron Laboratory, Michigan State University, East Lansing, Michigan 48824, USA}
\affiliation{Joint Institute for Nuclear Astrophysics, Michigan State University, East Lansing, Michigan 48824, USA}
\author{C.~J.~Prokop}
\affiliation{Department of Chemistry, Michigan State University, East Lansing, Michigan 48824, USA}
\affiliation{National Superconducting Cyclotron Laboratory, Michigan State University, East Lansing, Michigan 48824, USA}
\author{J.~Quaglia}
\affiliation{Department of Electrical Engineering, Michigan State University, East Lansing, Michigan 48824, USA}
\affiliation{Joint Institute for Nuclear Astrophysics, Michigan State University, East Lansing, Michigan 48824, USA}
\affiliation{National Superconducting Cyclotron Laboratory, Michigan State University, East Lansing, Michigan 48824, USA}
\author{S.~J.~Quinn}
\affiliation{Department of Physics and Astronomy, Michigan State University, East Lansing, Michigan 48824, USA}
\affiliation{National Superconducting Cyclotron Laboratory, Michigan State University, East Lansing, Michigan 48824, USA}
\affiliation{Joint Institute for Nuclear Astrophysics, Michigan State University, East Lansing, Michigan 48824, USA}
\author{J.~Sakstrup}
\affiliation{Department of Physics and Astronomy, Michigan State University, East Lansing, Michigan 48824, USA}
\affiliation{National Superconducting Cyclotron Laboratory, Michigan State University, East Lansing, Michigan 48824, USA}
\author{M.~Santia}
\affiliation{Department of Physics and Astronomy, Michigan State University, East Lansing, Michigan 48824, USA}
\affiliation{National Superconducting Cyclotron Laboratory, Michigan State University, East Lansing, Michigan 48824, USA}
\author{S.~B.~Schwartz}
\affiliation{Department of Physics and Astronomy, Michigan State University, East Lansing, Michigan 48824, USA}
\affiliation{National Superconducting Cyclotron Laboratory, Michigan State University, East Lansing, Michigan 48824, USA}
\affiliation{Geology and Physics Department, University of Southern Indiana, Evansville, Indiana 47712, USA}
\author{S.~Shanab}
\affiliation{Department of Physics and Astronomy, Michigan State University, East Lansing, Michigan 48824, USA}
\affiliation{National Superconducting Cyclotron Laboratory, Michigan State University, East Lansing, Michigan 48824, USA}
\author{A.~Simon}
\affiliation{National Superconducting Cyclotron Laboratory, Michigan State University, East Lansing, Michigan 48824, USA}
\affiliation{Department of Physics and Joint Institute for Nuclear Astrophysics, University of Notre Dame, Notre Dame, Indiana 46556, USA}
\author{A.~Spyrou}
\affiliation{Department of Physics and Astronomy, Michigan State University, East Lansing, Michigan 48824, USA}
\affiliation{National Superconducting Cyclotron Laboratory, Michigan State University, East Lansing, Michigan 48824, USA}
\affiliation{Joint Institute for Nuclear Astrophysics, Michigan State University, East Lansing, Michigan 48824, USA}
\author{E.~Thiagalingam}
\affiliation{Department of Physics and Astronomy, McMaster University, Hamilton, Ontario L8S 4M1, Canada}


\date{\today}

\begin{abstract}
\begin{description}
\item[Background] Measurements of $\beta$ decay  provide important nuclear structure information that can be used to probe isospin asymmetries and inform nuclear astrophysics studies.
\item[Purpose] To measure the $\beta$-delayed $\gamma$ decay of  $^{26}$P  and compare the results with previous experimental results and shell-model calculations.
\item[Method] A $^{26}$P fast beam produced using nuclear fragmentation was implanted into a planar germanium detector. Its $\beta$-delayed $\gamma$-ray emission was measured  with an array of 16 high-purity germanium detectors. Positrons emitted in the decay were detected in coincidence to reduce the background.
\item[Results] The absolute intensities of $^{26}$P $\beta$-delayed  $\gamma$-rays  were determined. A total of six new $\beta$-decay branches and 15 new $\gamma$-ray lines have been observed for the first time in $^{26}$P $\beta$-decay. A complete  $\beta$-decay scheme was built for the allowed transitions to bound excited states of $^{26}$Si. $ft$ values and Gamow-Teller strengths were also determined for these transitions and compared with shell model calculations and the mirror $\beta$-decay of  $^{26}$Na, revealing  significant mirror asymmetries.
\item[Conclusions]  A very good agreement with theoretical predictions based on the USDB shell model is observed. The  significant mirror asymmetry observed for the transition to the first excited state ($\delta=51(10)\%$) may be evidence for a proton halo in $^{26}$P. 
\end{description}
\end{abstract}


\maketitle

\section{Introduction\label{sec:intro}}
The detailed study of unstable nuclei was a major subject in nuclear physics during recent decades. $\beta$ decay measurements provide not only important information on the structure of the daughter and parent nuclei, but  can also be used to inform  nuclear astrophysics  studies and probe fundamental subatomic symmetries \cite{Hardy2015}. The link between experimental results and theory is given by the reduced transition probabilities, $ft$. Experimental $ft$ values involve three measured quantities: the half-life, $t_{1/2}$, the $Q$ value of the transition, which determines the statistical phase space factor $f$, and the branching ratio associated with that transition, $BR$.

In the standard $\mathcal{V\!\!-\!\!A}$ description of  $\beta$ decay, $ft$ values are related to the fundamental constants of the weak interaction and the matrix elements through this equation:

\begin{equation}
ft=\frac{\mathcal{K}}{g_V^2|\langle f|\tau|i\rangle|^2+g_A^2|\langle f|\sigma\tau|i\rangle|^2} ,
\label{eq:theo_ft}
\end{equation}

where $\mathcal{K}$ is a constant and $g_{V(A)}$ are the vector (axial) coupling constants of the  weak interaction; $\sigma$ and $\tau$ are the spin and isospin operators, respectively. Thus, a comparison of the experimental $ft$ values with the  theoretical ones obtained from the calculated matrix elements is a good test of the nuclear wave functions obtained with model calculations. However, to reproduce the $ft$ values measured experimentally, the axial-vector coupling constant $g_A$ involved in Gamow-Teller transitions has to be renormalized \cite{Wilkinson1973,WILKINSON1973_2}. The effective coupling constant $g'_A=q\times g_A$ is deduced empirically from experimental results and depends on the mass of the nucleus: The quenching factor is $q=0.820(15)$ in the $p$ shell \cite{Chou1993}, $q=0.77(2)$ in the $sd$ shell \cite{Wildenthal1983}, and $q=0.744(15)$ in the $pf$ shell \cite{Martinez1996}. Despite  several theoretical approaches attempting to reveal the origin of the quenching factor it is still not fully understood \cite{Brown2005}.

Another phenomenon which shows the limitations of our theoretical models is the so-called \emph{$\beta$-decay mirror asymmetry}. If we assume that the nuclear interaction is independent of  isospin, the theoretical description of $\beta$ decay is identical for the decay of a proton ($\beta^+$) or a neutron ($\beta^-$) inside a nucleus. Therefore, the $ft$ values corresponding to analog transitions should be identical. Any potential asymmetries are quantified by the asymmetry parameter $\delta=ft^+/ft^-1$, where the $ft^\pm$  refers to the $\beta^\pm$ decays in the mirror nuclei. The average value of this parameter is $(4.8\pm0.4)\%$ for $p$ and $sd$ shell nuclei  \cite{Thomas2004}. From a theoretical point of view the mirror asymmetry can have two origins: (a) the possible existence of exotic \emph{second-class currents} \cite{Wilkinson1970447,PhysRevLett.38.321,WilkinsonEPJ}, which are not allowed within the framework of the standard  $\mathcal{V\!\!-\!\!A}$ model of the weak interaction and (b) the breaking of the isospin symmetry between the initial or final nuclear states. Shell-model calculations were performed to test the isospin non-conserving part of the interaction in $\beta$ decay \cite{Smirnova2003441}. The main contribution to the mirror asymmetry from the nuclear structure was found to be from the difference in the matrix elements of the Gamow-Teller operator ($|\langle f|\sigma\tau|i\rangle|^2$), because of isospin mixing and/or differences in the radial wave functions.

Large mirror asymmetries  have been reported for transitions involving  halo states \cite{Tanihata2013}. For example, the asymmetry parameter for the  $A=17$ mirror decays $^{17}$Ne$\rightarrow^{17}$F and $^{17}$N$\rightarrow^{17}$O  to the first excited states of the respective daughters was  measured to be $\delta=(-55\pm9)\%$ and $\delta=(-60\pm1)\%$  in two independent experiments \cite{Borge1993, Ozawa1998}. This result was  interpreted as evidence for a proton halo in the first excited state of $^{17}$F assuming that the fraction of the $2s_{1/2}$ component of the valence nucleons remains the same in $^{17}$Ne and $^{17}$N. However, a different interpretation was also given in terms of charge dependent effects which increase  the $2s_{1/2}$ fraction in  $^{17}$Ne by about 50\% \cite{PhysRevC.55.R1633}. The latter result is also consistent with the high cross section obtained in the fragmentation of $^{17}$Ne \cite{Ozawa199418,Ozawa199663}, suggesting the existence of a halo in  $^{17}$Ne. More recently Kanungo \emph{et al.} reported the possiblity of a two-proton halo in $^{17}$Ne \cite{Kanungo200321}.
An extremely large mirror asymmetry was also observed in the mirror decay of $A=9$ isobars $^{9}$Li$\rightarrow^{9}$Be and $^{9}$C$\rightarrow^{9}$B. A value of $\delta=(340\pm70)\%$ was reported for the $^{9}$Li and $^{9}$C $\beta$-decay transitions to the 11.8 and 12.2 MeV levels of their respective daughters, which is the largest ever measured \cite{Bergmann2001427,Prezado2003}. Despite  the low experimental interaction cross sections measured with various targets in attempts to establish the halo nature of $^{9}$C \cite{Ozawa199663,Blank1997242}, recent results at intermediate energies \cite{Nishimura2006}, together with the anomalous magnetic moment \cite{Matsuta1995c153} and theoretical predictions \cite{0256-307X-27-9-092101,PhysRevC.52.3013,Gupta2002}, make $^{9}$C a proton halo candidate.
The potential relationship between large mirror asymmetries and halos is therefore clear. Precision measurements of mirror asymmetries in states involved in strong, isolated, $\beta$-decay transitions might provide a technique to probe halo nuclei that is complementary to total interaction cross section and momentum distribution measurements in knockout reactions \cite{Tanihata2013}.

Moreover, $\beta$ decay of proton-rich nuclei can be used for nuclear astrophysics studies. Large $Q_\beta$-values of  these nuclei not only allow the population of the bound excited states of the daughter, but also open  particle emission channels. Some of these levels correspond to astrophysically significant resonances which cannot be measured directly because of limited radioactive beam intensities. For example, the $^{25}\mathrm{Al}(p,\gamma)^{26}\mathrm{Si}$ reaction \cite{Wrede_2009} plays an important role in the abundance of the cosmic $\gamma$-ray emitter $^{26}\mathrm{Al}$.  The effect of this reaction is to  reduce the amount of ground state $^{26}\mathrm{Al}$, which is bypassed by the sequence $^{25}\mathrm{Al}(p,\gamma)^{26}\mathrm{Si}(\beta\nu)^{26m}\mathrm{Al}$, reducing therefore the intensity of the 1809-keV $\gamma$-ray line characteristic of the  $^{26}\mathrm{Al}$  $\beta$ decay \cite{Iliadis_96}. Thus it is important to constrain the $^{25}\mathrm{Al}(p,\gamma)^{26}\mathrm{Si}$ reaction rate.

 $^{26}$P is the most proton-rich bound phosphorus isotope. With a half-life of $43.7(6)$~ms and a $Q_{EC}$ value of $18258(90)$~keV \cite{Thomas2004}  the $\beta$ decay can be studied over a wide energy interval. $\beta$-delayed $\gamma$-rays  and protons from excited levels of $^{26}$Si  below and above  the proton  separation energy of $5513.8(5)$~keV \cite{AME2012} were observed directly in previous experiments \cite{Thomas2004,Cable_83,Cable_84} and, more recently, indirectly from the Doppler broadening of peaks in  the $\beta$-delayed  proton-$\gamma$ spectrum \cite{Schwartz2015}. The contribution of novae to the abundance  of $^{26}\mathrm{Al}$ in the galaxy was recently constrained by using  experimental data  on the $\beta$ decay of  $^{26}$P \cite{Bennett2013}.

In addition, $^{26}\mathrm{P}$ is a candidate to have a proton halo \cite{Brown1996,Ren1996,Gupta2002,Liang2009}. Phosphorus isotopes are the lightest nuclei expected to have a ground state with a dominant contribution of a $\pi s_{1/2}$ orbital. Low orbital angular momentum orbitals enhance the halo effect, because higher
$\ell$-values give rise to a confining centrifugal barrier. The low separation energy of  $^{26}$P (143(200)~keV \cite{AME2012}, 0(90)~keV\cite{Thomas2004}),   together with the narrow momentum distribution and enhanced cross section observed in proton-knockout reactions \cite{Navin1998} give some experimental evidence for the  existence of a proton halo in $^{26}$P. 

In this paper, we present a comprehensive summary of the $\beta$-delayed $\gamma$ decay of $^{26}$P
 measured at the National Superconducting Cyclotron Laboratory (NSCL) at Michigan State University during a fruitful experiment for which selected  results have already been reported in two separate shorter papers \cite{Bennett2013,Schwartz2015}. In the present work, the Gamow-Teller strength, $B(GT)$, and the experimental $ft$  values are compared to theoretical calculations and to the decay of the mirror nucleus  $^{26}$Na to investigate the  Gamow-Teller strength and mirror asymmetry, respectively. A potential relationship between  the mirror asymmetry and the existence of a proton halo in $^{26}$P is also discussed. Finally, in the last section, the calculated thermonuclear $^{25}$Al$(p,\gamma)^{26}$Si reaction rate, which was used in Ref. \cite{Bennett2013} to estimate the contribution of novae to the abundance of galactic $^{26}$Al, is tabulated for completeness.
 

\section{Experimental procedure\label{sec:experiment}}

\begin{figure}[t]
\centering
\includegraphics[width=0.45\textwidth]{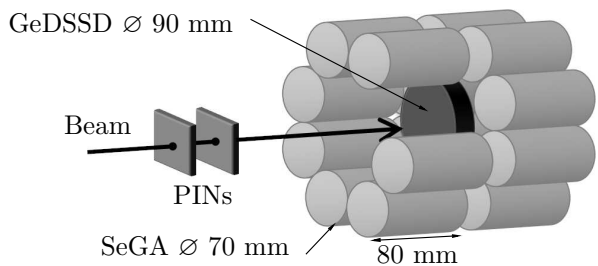}%
\caption{\label{fig:Setup} Schematic view of the experimental setup. The thick arrow indicates the beam direction. One of the  16 SeGA detectors was removed to show the placement of the GeDSSD.}
 \end{figure}

The experiment was carried out at the National Superconducting Cyclotron Laboratory (NSCL). A 150~MeV/u 75~pnA primary beam of $^{36}\mathrm{Ar}$ was delivered from the Coupled Cyclotron Facility and impinged upon a 1.55~g/cm$^2$ Be target. The $^{26}\mathrm{P}$ ions were in-flight separated from other fragmentation products according to their magnetic rigidity by the A1900 fragment separator \cite{Morrissey200390}. The Radio-Frequency Fragment Separator (RFFS) \cite{Bazin2009314} provided a further increase in beam purity before the beam was implanted into a 9-cm diameter, 1-cm thickness planar germanium double-sided strip detector (GeDSSD) \cite{Larson201359}. To detect signals produced by both the implanted ions and the $\beta$ particles emitted during the decay, the GeDSSD was connected to two parallel amplification chains. This allowed the different amounts of energy deposited in implantations (low gain) and decays (high gain) to be detected in the GeDSSD. The GeDSSD was surrounded by the high purity germanium detector array SeGA \cite{Mueller2001492} in its barrel configuration which was used to  measure the $\beta$-delayed $\gamma$ rays (see Fig.\ref{fig:Setup}).

\begin{figure}[b]
 \includegraphics[trim=0cm 0mm 0cm 1cm, clip=true,width=.5\textwidth]{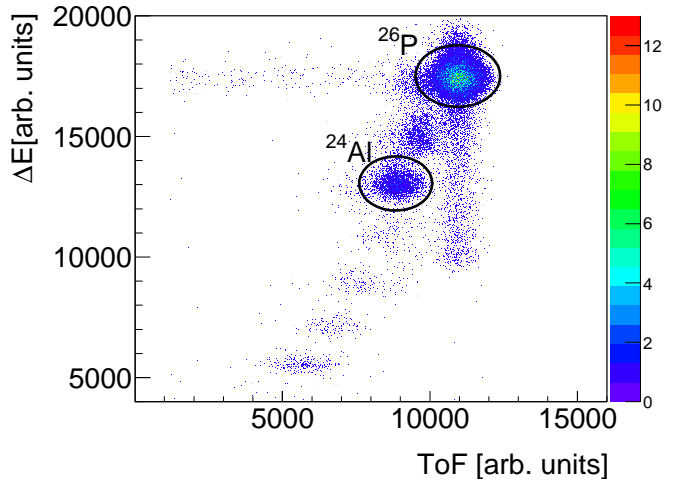}%
 \caption{\label{fig:PID} Particle identification plot obtained for a selection of runs during the early portion of the experiment, before the beam tune was fully optimized. The energy loss was obtained from one of the PIN detectors and the time of flight between the same detector and the scintillator placed at the focal plane of the A1900 separator. A low-gain energy signal in the GeDSSD condition was used. The color scale corresponds to the number of ions.}
 \end{figure}

  The identification of the incoming beam ions was accomplished using   time-of-flight and energy loss signals. The energy loss signals were provided by a pair of  silicon PIN detectors  placed slightly upstream of the decay station. The time of flight was measured between one of these PINs and a plastic scintillator placed 25~m upstream, at the A1900 focal plane. Figure  \ref{fig:PID} shows a two-dimensional cluster  plot of the  energy loss versus the time of flight for the incoming  beam taken prior to a re-tune that improved the beam purity substantially for the majority of the experiment. A coincidence condition requiring a low-gain signal in the GeDSSD was applied to ensure the ions were implanted in the detector. It shows that the main contaminant in our beam was the radioactive isotone $^{24}\mathrm{Al}$ ($\sim$13\%). During the early portion of the experiment, a small component of $^{25}\mathrm{Si}$ was also present in the beam. We estimated its ratio and it was on average 2.1\%, but this value was diluted to 0.5\% after incorporating the data acquired after the re-tune.  Small traces of lighter isotones like $^{22}\mathrm{Na}$ and $^{20}\mathrm{F}$ were also present ($\sim$2.5\%). The total secondary beam rate was on average 80~ions/s and the  overall purity of the implanted  beam was 84\%. This value of the beam purity differs from the previous reported values in Ref. \cite{Bennett2013}, in which the implant condition was not applied The  $^{26}\mathrm{P}$ component was composed of the ground state and the known  164.4(1)~keV isomeric state \cite{Nishimura2014,DPL2016}. Because of the short half-life of the isomer [120(9)~ns] \cite{Nishimura2014} and the fact that it decays completely to the ground state of $^{26}\mathrm{P}$, our $\beta$-decay measurements were not affected by it.

The data were collected event-by-event using the NSCL digital acquisition system \cite{Prokop2014}. Each channel provided its own time-stamp signal, which allowed coincidence gates to be built between the different detectors. To select $\beta$-$\gamma$ coincidence events, the high-gain energy signals from the GeDSSD were used to indicate that a $\beta$ decay occurred. The subsequent $\gamma$ rays emitted from excited states of the daughter nuclei were selected by setting a 1.5-$\mu$s coincidence window. The 16  spectra obtained by each of the elements of SeGA were then added together after they were gain  matched  run-by-run to account for  possible gain drifts during the course of the experiment.

\begin{figure*}
\centering
 \includegraphics[trim=2cm 5mm 5mm 2mm, clip=true,width=.85\textwidth]{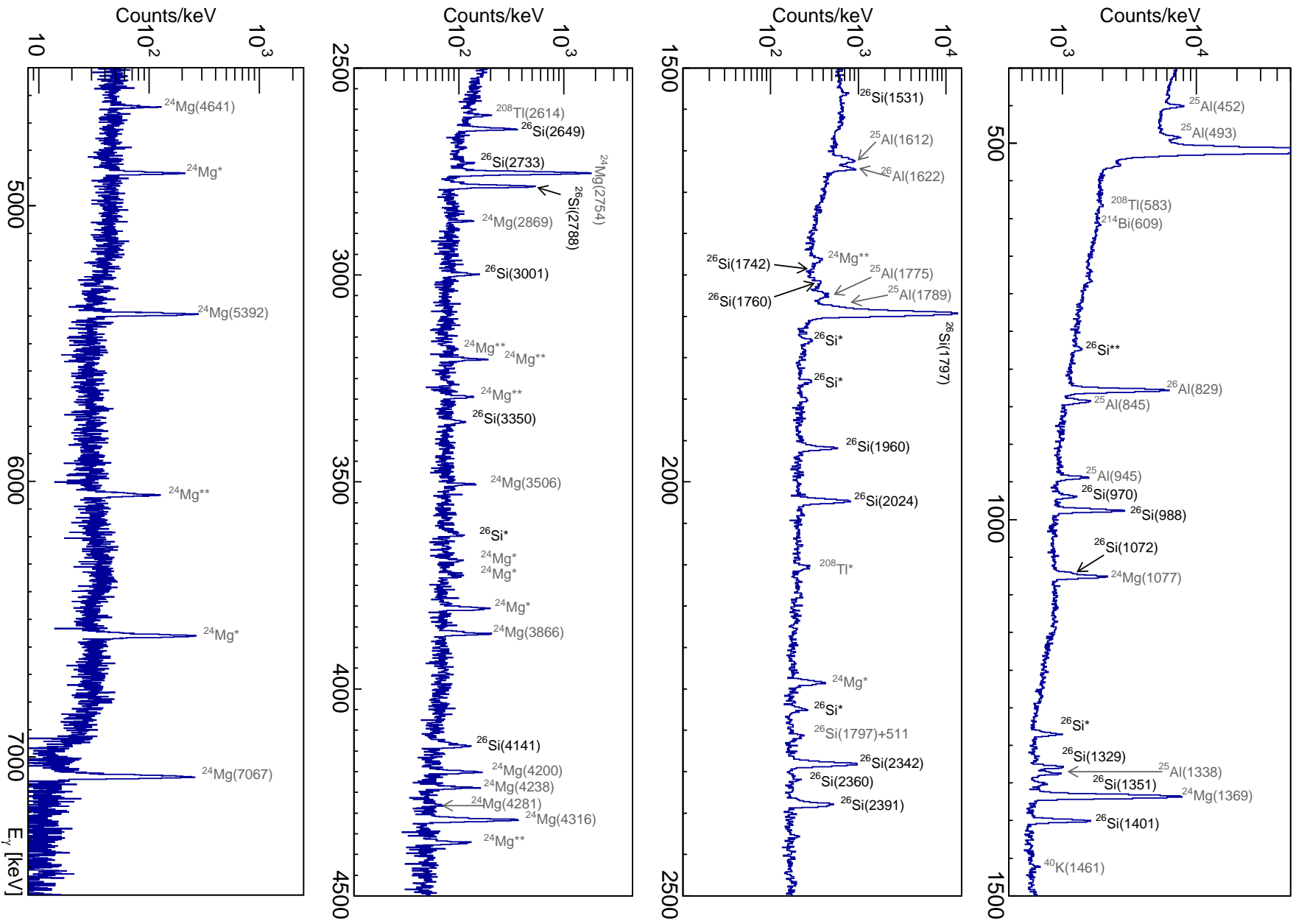}%
 \caption{\label{fig:spec} $\gamma$-ray spectrum observed by the SeGA array in coincidence with a $\beta$ particle in the GeDSSD.  Photopeaks have been labeled by the emitting nucleus and its energy rounded to the closest keV integer. Peaks labeled with one (two) asterisks correspond to single (double) escape peaks.}
 \end{figure*}

\section{Data Analysis and Experimental Results \label{sec:data}}

As   mentioned in  Sec. \ref{sec:intro}, the data presented in this paper are from the same experiment described in Refs. \cite{Bennett2013,Schwartz2015}, but independent sorting and analysis routines were developed and employed. The values extracted are therefore slightly different, but consistent within uncertainties. New values derived in the present work are not intended to supersede those from Refs. \cite{Bennett2013,Schwartz2015}, but rather to complement them. In this section, the analysis procedure is described in detail and the experimental results are presented.

Figure  \ref{fig:spec} shows the cumulative $\gamma$-ray spectrum observed in all the detectors of the SeGA array in coincidence with a $\beta$-decay signal in the GeDSSD. We have identified 48 photopeaks, of which 30 are directly related to the decay of $^{26}$P. Most of the other peaks were assigned to the $\beta$ decay of the main contaminant of the beam,  $^{24}$Al. Peaks in the spectrum have been labeled by the $\gamma$-ray emitting nuclide. Twenty-two of the peaks correspond to $^{26}$Si, while eight of them  correspond to  $\beta$-delayed proton decays to  excited states of $^{25}$Al followed by $\gamma$-ray emission. In this work we will focus on the decay to  levels of $^{26}$Si as the $^{25}$Al levels have already been discussed in Ref. \cite{Schwartz2015}.


\begin{figure}
 \includegraphics[trim=5mm 1cm 1cm 0mm, clip=true,width=.5\textwidth]{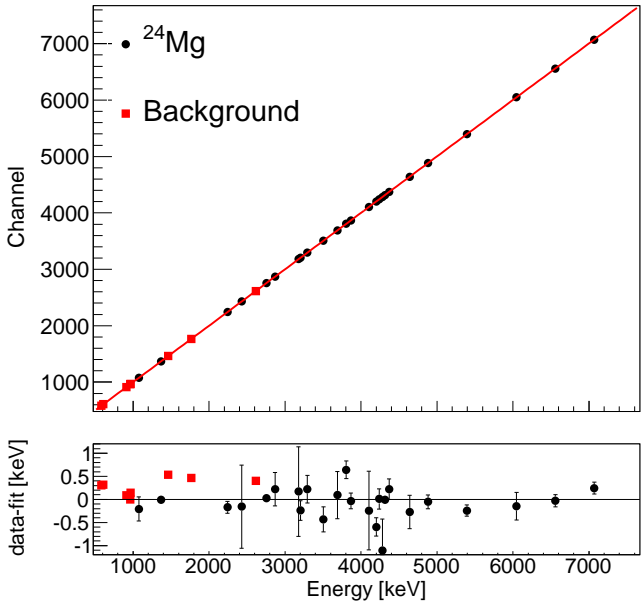}%
 \caption{\label{fig:calibr}(Upper panel) Energy calibration of SeGA $\gamma$-ray spectra using the  $\beta$-delayed $\gamma$ rays emitted by $^{24}\mathrm{Al}$. The solid line is the result of a second degree polynomial fit. Energies and uncertainties are taken from \cite{Firestone20072319}. (Lower panel) Residuals of the calibration points with respect to the calibration line.}
 \end{figure}

\subsection{$\bm{\gamma}$-ray Energy Calibration}

The energies of the $\gamma$ rays emitted during the experiment were determined from a calibration of the SeGA array. As mentioned in  Sect. \ref{sec:experiment} and in Refs. \cite{Schwartz2015,Bennett2013} a gain-matching procedure was performed to align all the signals coming from the 16 detectors comprising the array. This alignment was done with the strongest background peaks, namely the 1460.8-keV line (from the  $^{40}\mathrm{K}$ decay) and the 2614.5-keV one (from the $^{208}\mathrm{Tl}$ decay). The gain-matched cumulative spectrum was then absolutely calibrated \emph{in situ} using the well-known energies of the $^{24}$Al  $\beta$-delayed $\gamma$ rays emitted by $^{24}\mathrm{Mg}$, which cover a wide range in energy from 511~keV to almost 10~MeV \cite{Firestone20072319}. To account for possible non-linearities in the response of the germanium detectors, a second degree polynomial fit was used as a calibration function.  Results of the calibration are shown in Fig.~\ref{fig:calibr}. The standard deviation for this fit is 0.3~keV, which includes the literature uncertainties associated with  the energies of $^{24}\mathrm{Mg}$. The systematic uncertainty was estimated from the residuals of room background peaks not included in the fit. The lower panel of Fig.~\ref{fig:calibr} shows that these deviations are below 0.6~keV, with an average of 0.2~keV. Based on this, the systematic uncertainty was estimated to be 0.3~keV.

\begin{figure}
 \includegraphics[trim=5mm 1cm 1cm 0mm, clip=true,width=.5\textwidth]{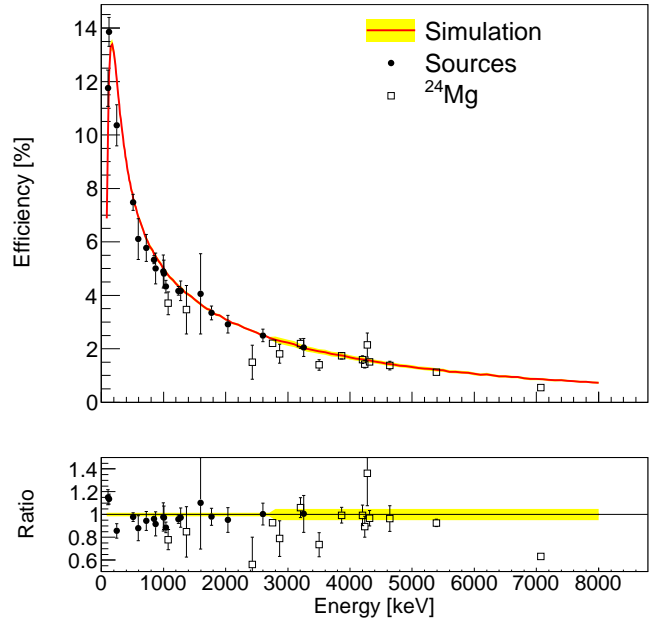}%
 \caption{\label{fig:eff}SeGA photopeak efficiency. (Top panel) Results of a {\sc Geant4} simulation [solid line (red)] compared to the efficiency measured with absolutely calibrated sources (black circles) and the known $^{24}\mathrm{Mg}$ lines (empty squares). The simulation and the $^{24}\mathrm{Mg}$ data have been scaled to match the source measurements. (Bottom panel) Ratio between the simulation and the experimental data. The shaded area (yellow) shows the adopted uncertainties.}
 \end{figure}

\subsection{Efficiencies}

\subsubsection{$\beta$-particle Efficiency \label{sec:betaeff}}
The $\beta$-particle detection efficiency of the GeDSSD can be determined  by taking the ratio between the number of counts under a certain photopeak in the $\beta$-gated $\gamma$-ray singles spectrum and the ungated one. In principle, the $\beta$ efficiency depends on  $Q_\beta$. To investigate this effect, we calculated the ratios between the gated and the ungated spectra for all the $^{24}\mathrm{Mg}$ peaks, which have different combinations of $Q_\beta$, and found it to be independent of the end-point energy of the $\beta$ particles, with an average ratio of $\varepsilon_\beta(^{24}\mathrm{Mg})=(38.6\pm0.9)\%$. Because of the different implantation depths for $^{24}\mathrm{Al}$ and $^{26}\mathrm{P}$ ($^{24}\mathrm{Al}$ barely penetrates into the GeDSSD), we also calculated the gated to ungated ratios of the strongest peaks of $^{26}\mathrm{Si}$ (1797~keV) and its daughter $^{26}\mathrm{Al}$ (829~keV) obtaining a constant, average, value for the efficiency of $\varepsilon_\beta=(65.2\pm0.7)\%$. The singular value for $^{26}\mathrm{Si}$ and $^{26}\mathrm{Al}$ is explained by their common decay point in the GeDSSD.

\subsubsection{$\gamma$-ray  Efficiency}

To obtain  precise measurements of the $\gamma$-ray intensities, we determined  the photopeak efficiency  of  SeGA. The photopeak efficiency was  studied over a wide energy range between 400 keV and 8 MeV. The results of a {\sc Geant4} \cite{Agostinelli2003250} Monte-Carlo simulation were compared with the relative intensities of the well-known $^{24}\mathrm{Mg}$ lines used also in the energy calibration. The high energy lines of this beam contaminant made it possible to benchmark the simulation for energies higher than  with standard sources. In addition,  the comparison of the simulation to data taken offline with absolutely-calibrated $^{154,155}\mathrm{Eu}$ and $^{56}\mathrm{Co}$ sources allowed us to scale the simulation to determine the  efficiency at any energy. The scaling factor was 0.91. The statistical uncertainty of this scaling factor was inflated by  a scaling factor of $\sqrt{\chi^2/\nu}$  yielding an uncertainty of 1.5\%, which was propagated into the efficiency. The magnitude of this  factor is consistent with {\sc Geant4} simulations of the  scatter associated with coincidence summing effects \cite{Semkow1990}.  Figure \ref{fig:eff} shows the adopted efficiency curve compared to the source data, and the  $^{24}\mathrm{Mg}$ peak intensities. The accuracy of this photopeak efficiency was estimated to be $\delta\varepsilon/\varepsilon=1.5\%$ for energies below 2800~keV and 5\% above that energy.

\subsection{$\bm{\gamma}$-ray intensities \label{subsec:intensities}}
\begin{figure}[t]
 \includegraphics[trim=0mm 1cm 1cm 0mm, clip=true,width=.5\textwidth]{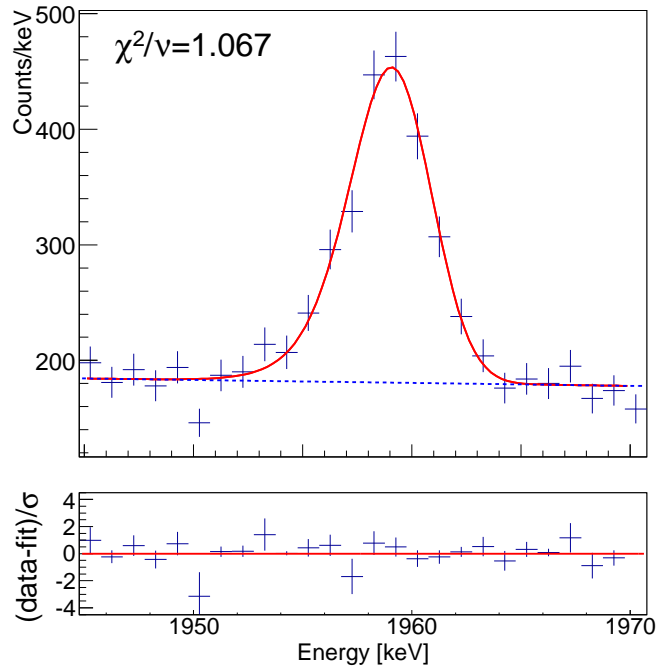}%
 \caption{\label{fig:fit} (Top panel) Example of a typical fit to the 1960-keV peak, using the function of Eq. (\ref{eq:EMG}). The dashed line corresponds to the background component of the fit. (Bottom panel) Residuals of the fit in terms of the standard deviation $\sigma$.}
 \end{figure}

The intensities of the $\gamma$ rays emitted  in the $\beta$ decay of $^{26}\mathrm{P}$ were obtained from the areas of the photopeaks shown in the spectrum of Fig. \ref{fig:spec}. We used an exponentially modified Gaussian (EMG) function to describe the peak shape together with a linear function to model the local background:

\begin{equation}
F=B+\frac{N}{2\tau}e^{\frac{1}{2\tau}\left (2\mu+\frac{\sigma^2}{\tau} -2x \right )}\mathrm{erfc}\!\left[\frac{ \sigma^2+\tau(\mu-x)}{\sqrt{2}\sigma\tau}\right ],
\label{eq:EMG}
\end{equation}

where $B$ is a linear background, $N$ is the area below the curve, $\mu$ and $\sigma$ are the centroid and the width of the Gaussian, respectively, and $\tau$ is the decay constant of the exponential; erfc is the complementary error function.
 The parameters describing the width of the Gaussian ($\sigma$) and the exponential constant ($\tau$) were determined by fitting narrow isolated peaks at various energies. The centroids and the areas below the peaks were obtained from the fits. When multiple peaks were very close, a multi-peak fitting function was applied using  the same values for the $\tau$ and $\sigma$ parameters for all the peaks in the region. In general the fits were very good, with reduced chi-squared ($\chi^2/\nu$) close to  unity. In those cases where  $\chi^2/\nu$ was bigger than one, the statistical uncertainties were inflated by multiplying them by $\sqrt{\chi^2/\nu}$. Fig. \ref{fig:fit} shows an example of the fit to the 1960-keV peak.

\subsubsection{Absolute normalization}
The total number of $^{26}\mathrm{P}$ ions  implanted and subsequently decaying in the GeDSSD is, in principle, needed to obtain an absolute normalization of the $\gamma$-ray intensities, and hence the $\beta$ branchings of $^{26}\mathrm{Si}$ levels. The number of $\gamma$ rays observed  at energy $E$ is: 
\begin{equation}
N_\gamma(E)=N_0 \times \varepsilon_{\gamma}(E)\times \varepsilon_{\beta}(E) \times I_{\gamma}(E)
\label{eq:abs_intensity}
\end{equation}

\begin{table}
 \caption{\label{tab:levels} Data on $^{26}$Si energy levels and  $^{26}\mathrm{P}(\beta\gamma)$ decay. A total of 12 levels and 22 $\gamma$ rays have been identified. The first column shows the level energies  obtained from the laboratory $\gamma$-ray energies shown in the fifth column and include the nuclear recoil correction factor. The second column shows the $\beta$-branches. The third and fourth columns show the spin and parity of the initial and final state, respectively. The last column corresponds to the absolute intensities of the $\gamma$ rays.}
 \begin{ruledtabular}

\begin{tabular}[c]{d d c c d d}
\multicolumn{1}{c}{$E_x$ (keV)} & \multicolumn{1}{c}{$\beta$-Branch (\%)} & $_iJ_n^\pi$ & $_fJ_n^\pi$ & \multicolumn{1}{c}{ $E_\gamma$ (keV)} & \multicolumn{1}{c}{$I$ (\%)} \\
\hline
1797.1(3)   & \multicolumn{1}{c}{ 41(3) }  & $2_1^+$    & $0_1^+$     &  1797.1(3)       &\multicolumn{1}{c}{ 58(3)}    \\       
2786.4(3)   & \textless0.39\footnote[3]{95\% confidence level.}       & $2_2^+$    & $2_1^+$     &  989.0(3)     & 5.7(3)   \\
            &                     &            & $0_1^+$     &  2786.5(4)\footnote[1]{Transition never observed in $^{26}$P $\beta$ decay}       & 3.4(2)   \\
3756.8(3)   & 1.9(2)              & $3_1^+$    & $2_2^+$     &  970.3(3)        & 1.15(9)   \\
            &                     &            & $2_1^+$     &  1959.8(4)       & 1.7(1)   \\
4138.6(4)   & 6.2(4)              & $2_3^+$    & $2_2^+$     &  1352.2(4)\footnotemark[1]       & 0.48(7)   \\
            &                     &            & $2_1^+$     &  2341.2(4)       & 4.7(3)   \\
            &                     &            & $0_1^+$     &  4138.0(5)\footnotemark[1]       & 1.0(1)   \\
4187.6(4)   & 4.4(3)              & $3_2^+$    & $2_2^+$     &  1401.3(3)       & 3.8(2)   \\
            &                     &            & $2_1^+$     &  2390.1(4)\footnotemark[1]       & 2.2(1)   \\
4445.1(4)   & 0.8(2)\footnotemark[1]              & $4_1^+$    & $2_2^+$     &  \multicolumn{1}{l}{ 1660(2)\footnotemark[1]}         & 0.08(6)   \\
            &                     &            & $2_1^+$     &  2647.7(5)\footnotemark[1]       & 1.7(1)   \\
4796.4(5)   & 0.56(9)\footnotemark[1]             & $4_2^+$    & $2_2^+$     &  2999.1(5)\footnotemark[1]        & 0.56(9)   \\
4810.4(4)   & 3.1(2)\footnotemark[1]              & $2_4^+$    & $2_2^+$     &  2023.9(3)\footnotemark[1]        & 3.1(2)   \\
5146.5(6)   & 0.18(5)\footnotemark[1]              & $2_5^+$    & $2_2^+$     &  2360.0(6)\footnotemark[1]        & 0.18(5)   \\
5288.9(4)   & 0.76(7)\footnotemark[1]             &  $4_3^+$   &  $4_1^+$    &  842.9(3)\footnotemark[1]         & 0.33(7)  \\
            &                     &            & $3_1^+$     &  1532.1(5)\footnotemark[1]        & 0.43(7)   \\
            &                     &            & $2_1^+$     & \multicolumn{1}{l}{  3491\footnote[2]{Not observed.}}   &\textless0.12\footnotemark[3]   \\
5517.3(3)   & 2.7(2)\footnotemark[1]              &  $4_4^+$   &  $4_1^+$    &  1072.1(5)\footnotemark[1]         & 0.69(9)  \\
            &                     &            & $3_2^+$     &  1329.9(3)\footnotemark[1]         & 1.4(1)   \\
            &                     &            & $3_1^+$     &  1759.7(5)\footnotemark[1]         & 0.47(6)   \\
            &                     &            & $2_2^+$     &  2729.9(5)\footnotemark[1]         & 0.29(5)   \\
5929.3(6)   & 0.15(5) \footnote[4]{Only the $\gamma$ branch has been measured.}             &  $3_3^+$   &  $3_2^+$    &  1741.7(9)         & 0.15(5)   \\
 \end{tabular}
 \end{ruledtabular}
 \end{table}

where  $N_0$  is the total number of ions decaying, $\varepsilon_{\gamma(\beta)}$ are the efficiencies to detect $\gamma$ rays ($\beta$ particles), and $I_{\gamma}$ is the absolute $\gamma$-ray intensity. To circumvent the uncertainty associated with the  total number of ions decaying,  we  used  the ratio of the number of $\beta$ decays of $^{26}\mathrm{P}$ to its daughter $^{26}\mathrm{Si}$ [$61(2)\%$] \cite{Thomas2004}, and the absolute intensity of the 829-keV $\gamma$-rays emitted in the $\beta$ decay of  $^{26}\mathrm{Si}$, $[21.9(5)\%]$ \cite{Endt19981}, to calculate the intensity of the 1797-keV line, which is the most intense $\gamma$ ray emitted in the decay of $^{26}\mathrm{P}$ (see Table \ref{tab:levels}). To do so, we applied Eq. (\ref{eq:abs_intensity}) to these two $\gamma$ rays :

\begin{equation}
\label{eq:intensity_Al}
N_\gamma(829) = N_{^{26}\mathrm{Si}} \varepsilon_{\gamma}(829) \varepsilon_{\beta}(829)  I_{\gamma}(829)
\end{equation}
\begin{equation}
N_\gamma(1797) = N_{^{26}\mathrm{P}} \varepsilon_{\gamma}(1797) \varepsilon_{\beta}(1797)  I_{\gamma}(1797)
\label{eq:intensity_Si}
\end{equation}

By taking the ratio between Eqs. (\ref{eq:intensity_Al}) and (\ref{eq:intensity_Si}), the only unknown is the intensity of the 1797-keV $\gamma$ ray, because the $\beta$ efficiencies can be obtained from the  $\beta$-gated to ungated ratios discussed in Sec. \ref{sec:experiment}. The value obtained for the intensity of the 1797-keV $\gamma$ ray is thus 58(3)\%, which is in agreement with the value 52(11)\% reported in Ref. \cite{Thomas2004} and more precise. The rest of the  $\gamma$-ray intensities  were determined with respect to this value by employing the efficiency curve and they are  presented in Table \ref{tab:levels}. We also report an upper limit on the intensity of one $\gamma$ ray which was expected to be near the theshold of our sensitivity given the intensity predicted by theory.

\subsection{$\bm{\beta}$-$\bm{\gamma}$-$\bm{\gamma}$ coincidences \label{subsec:coincidences}}
\begin{figure}[t]
 \includegraphics[trim=0mm 5mm 1cm 0mm, clip=true,width=.5\textwidth]{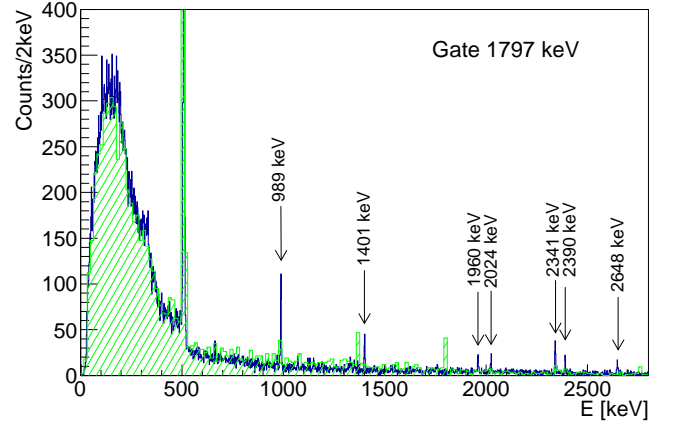}%
 \caption{\label{fig:Coincidences2} (Color online) $\beta$-$\gamma$-$\gamma$ coincidence  spectrum gating on   the 1797 keV $\gamma$-rays (blue). The hatched histogram (green) shows  coincidences with continuum background in a relatively broad region above the peak gate.  The background bins are 16 keV wide and are normalized to the expected background per 2 keV from random coincidences. The strongest peaks corresponding to $\gamma$ rays emitted in coincidence are indicated.}
 \end{figure}

 The 16-fold granularity of SeGA  allowed us  to obtain $\beta$-$\gamma$-$\gamma$ coincidence spectra,  which helped to interpret the $^{26}\mathrm{P}$ decay scheme. Fig. \ref{fig:Coincidences2} shows the gamma coincident spectrum gated on the 1797-keV peak, where we can see several peaks corresponding to $\gamma$ rays detected  in coincidence. To estimate the background from random coincidences, we have created another histogram gated on the background close to the peak and normalized  to the number of counts within the gated regions. At some energies the background estimate is too high. This is because of a contribution from real $\gamma$-$\gamma$ coincidences involving Compton background, which should not be normalized according to the random assumption.

\begin{figure*}
\centering
 \includegraphics[trim=0mm 2mm 1cm 9mm, clip=true,width=.75\textwidth]{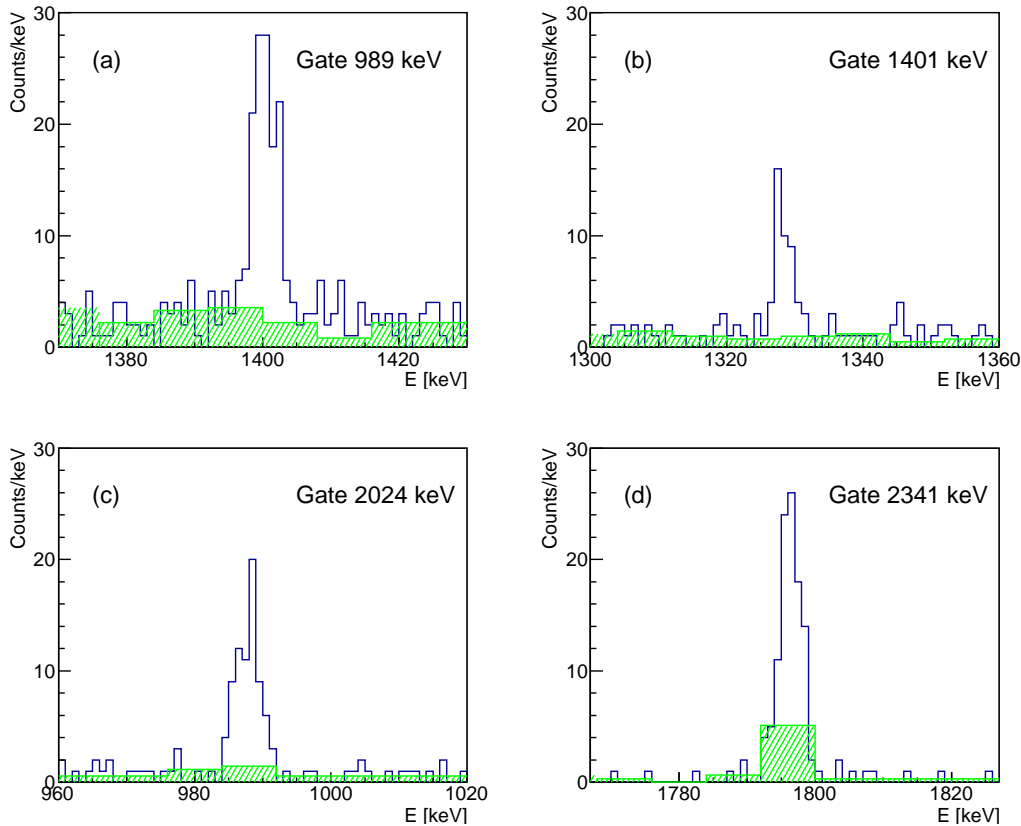}%
 \caption{\label{fig:Coincidences} Selected sample of  $\beta$-$\gamma$-$\gamma$ coincidence  peaks gating on different $\gamma$ rays: (a) 989~keV, (b) 1401~keV, (c) 2024~keV, and (d)  2341~keV. The hatched histogram shows  normalized coincidences with continuum background in a relatively broad region above the peak gates. The background bins are 8-keV wide  and are normalized to the expected background per keV.}
 \end{figure*}

   Fig. \ref{fig:Coincidences} presents a sample of  peaks observed in  coincidence when gating on  some other intense $\gamma$ rays observed. From this sample we can see that the  coincidence technique helps to  cross-check the decay scheme. For example Fig. \ref{fig:Coincidences}(a) shows clearly that the 1401-keV $\gamma$ ray is emitted in coincidence with the 989-keV $\gamma$ ray, indicating that the former $\gamma$ ray comes from a  higher-lying level. In the same way, we can see in  Fig. \ref{fig:Coincidences}(b)  that the 1330-keV $\gamma$-ray  is emitted from a level higher than the 4187-keV level. From the gated spectra, some information can  also be extracted from the missing peaks. As  Fig. \ref{fig:Coincidences}(c) shows, by gating on the 2024-keV $\gamma$ ray the 970-keV peak disappears, displaying only the 989-keV peak, which means that the 970-keV $\gamma$ ray comes from a level which is  not connected with these two levels by any $\gamma$-ray cascade. Fig. \ref{fig:Coincidences}(d) shows clearly the coincidence between the  $\gamma$ ray emitted from the first  $2^+$ state at 1797~keV to the ground state of $^{26}$Si and the  2341-keV $\gamma$ ray from the third $2^+$ state to the first excited state.

 These coincidence procedures were systematically analyzed for all possible combinations of $\gamma$ rays and the results are summarized in Table \ref{tab:coincidence} in the form of a 2D matrix, where a checkmark (\checkmark) means the $\gamma$ rays were detected in coincidence. The condition for a $\gamma$ ray to be listed in coincidence with another is for it to be at least 3$\sigma$ above the estimated random-coincidence background. It is  worth noting  that this background estimate is somewhat conservative, therefore  the significance of some of the peaks is underestimated.

\begin{table*}
 \caption{\label{tab:coincidence}  Coincidence matrix of all the $\gamma$ rays measured in the $\beta$ decay of $^{26}\mathrm{P}$. The first column corresponds to the  $\gamma$-ray energy on which the gate is set. The following columns indicate the $\gamma$ rays observed in the gated spectrum. $\gamma$ rays observed in coincidence are indicated with a checkmark (\checkmark) if the detection is larger than $3\sigma$ above background. $\gamma$-ray energies have been rounded to the closest integer and are given in keV.}
 \begin{ruledtabular}

\begin{tabular}[c]{ccccccccccccccccccccccc}
	&843	&  970  &   989  &   1072  &  1330  &  1352  &  1401 &   1532&    1660  &  1742  &  1760  &  1797 &   1960  &  2024  &  2341 &   2360 & 2390	&  2648  &  2730  &  2787  &  2999  &  4138 \\
\hline
843	&	&  -	&   \checkmark	 &   -    &  -    &  -    &  -   &	 -  &	  -	&  -	 &  -    &  -   &   \checkmark     &  -    &  -   &	 -   &   -	&  \checkmark	 &  -    &  \checkmark     &  -    &  -   \\
970	&-	&	&   \checkmark	 &   \checkmark     &  -    &  -    &  -   &	 -   &	  -	&  -	 &  \checkmark	  &  \checkmark    &   -    &  -    &  -   &	 -   &   -	&  -	 &  -    &  -    &  -    &  -   \\
989	&\checkmark&  \checkmark&	 &   \checkmark     &  \checkmark  &  \checkmark  &  \checkmark    &\checkmark   &\checkmark	&  -	 & \checkmark& \checkmark &   -  &\checkmark&  - &\checkmark &   -	&  -	 &  \checkmark&  \checkmark&  -   &  -   \\
1072	&-	&  -	&   \checkmark	 &	   &  -    &  -    &  \checkmark    &	 -  &	  -	&  -	 &  -    &  \checkmark    &   -    &  -    &  -   &	 -   &   -	&  \checkmark	 &  -    &  -    &  \checkmark     &  -   \\
1330	&-	&  -	&   \checkmark	 &   -    &	    &  -    &  \checkmark    &	 -  &	  -	&  -	 &  -    &  \checkmark    &   \checkmark     &  -    &  -   &	 -   &   -	&  -	 &  -    &  -    &  -    &  -   \\
1352	&-	&  -	&   \checkmark	 &   -    &  -    &	     &  -   &	 -  &	  -	&  -	 &  -    &  \checkmark    &   -    &  -    &  -   &	 -   &   -	&  -	 &  -    &  \checkmark     &  -    &  -   \\
1401	&\checkmark	&  -	& \checkmark& \checkmark & \checkmark &  -  & & -  &	  -	& \checkmark\footnote[1]{Not 3$\sigma$, but 99.6\% C.L.} &  -    & \checkmark & \checkmark &  -    &  -   &	 -   &   -	& -&  -  &\checkmark  &  -    &  -   \\
1532	&\checkmark	&  \checkmark	&   \checkmark	 &   --   &  -    &  -    &  -   &	     &	  -	&  -	 &  -    &  \checkmark    &   \checkmark     &  -    &  -   &	 -   &   -	&  -	 &  -    &  -    &  -    &  -   \\
1660	&\checkmark	&  -	&   -   &   \checkmark     &  -    &  -    &  -   &	 -  &	 	&  -	 &  -    &  \checkmark    &   -    &  -    &  -   &	 -   &   -	&  -	 &  -    &  \checkmark     &  -    &  -   \\
1742	&-	&  -	&   -	 &   -    &  -    &  -    &  \checkmark    &	 -  &	  -	&	 &  -    &  -    &   -    &  -    &  -   &	 -   &   -	&  -	 &  -    &  -     &  -    &  -   \\
1760	&-	&  \checkmark	&   \checkmark	 &   -    &  -    &  -    &  -   &	 -  &	  -	&  -	 &	  &   \checkmark   &   \checkmark     &  -    &  -   &	 -   &   -	&  -	 &  -    &  -    &  -    &  -   \\
1797	&-	&  -	&   \checkmark	 &   -     &  -     &  -  &  \checkmark    &	 -   &	  -	&  -	 &  -	  &	  &   \checkmark     &  \checkmark     &  \checkmark   & -   &   \checkmark	&  \checkmark	 &  -	  &  -    &  -     &  -   \\
1960	&\checkmark	&  -	&   -   &   -    &  \checkmark     &  -    &  -   &	 \checkmark   &	  -	&  -	 &  \checkmark	  &  \checkmark    &	    &  -    &  -   &	 -   &   -	&  -	 &  -    &  -    &  -    &  -   \\
2024	&\checkmark	&  -	&   \checkmark	 &   -    &  -    &  -    &  -   &	 -  &	  -	&  -	 &  -    &  \checkmark    &   -    &	     &  -   &	 -   &   -	&  -	 &  -    &  \checkmark     &  -    &  -   \\
2341	&-	&  -	&   -   &   -    &  \checkmark     &  -    &  -   &	 -  &	  -	&  -	 &  -    &  \checkmark    &   -    &  -    &	     &	 -   &   -	&  -	 &  -    &  -    &  -    &  -   \\
2360	&-	&  -	&   -   &   -    &  -    &  -    &  -   &	 -  &	  -	&  -	 &  -    &  \checkmark    &   -    &  -    &  -   &	      &   -	&  -	 &  -    &  \checkmark     &  -    &  -   \\
2390	&-	&  -	&   -   &   -    &  \checkmark     &  -    &  -   &	 -  &	  -	&  -	 &  -    &  \checkmark    &   -    &  -    &  -   &	 -   &  	&  -	 &  -    &  -    &  -    &  -   \\
2648	&\checkmark	&  -	&   -   &   \checkmark     &  -    &  -    &  -   &	 -  &	  -	&  -	 &  -    &  \checkmark    &   -    &  -    &  -   &	 -   &   -	&	 &  -    &  -    &  -    &  -   \\
2730	&-	&  -	&   \checkmark	 &   -    &  -    &  -    &  -   &	 -  &	  -	&  -	 &  -    &  \checkmark    &   -    &  -    &  -   &	 -   &   -	&  -	 &	  &  -    &  -    &  -   \\
2787	&\checkmark	&  \checkmark	&   -   &   -    &  \checkmark     &  \checkmark     &  \checkmark    &	 -  &	  -	&  -	 &  \checkmark	  &  -   &   -    &  \checkmark     &  -   &	 -   &   -	&  -	 &  -    &	   &  -    &  -   \\
2999	&-	&  -	&   -   &   -    &  -    &  -    &  -   &	 -  &	  -	&  -	 &  -    &  \checkmark    &   -    &  -    &  -   &	 -   &   -	&  -	 &  -    &  -    &	    &  -   \\
4138	&-	&  -	&   -   &   -    &  -    &  -    &  -   &	 -  &	  -	&  -	 &  -    &  -   &   -    &  -    &  -   &	 -   &   -	&  -	 &  -    &  -    &  -    &	    \\

 \end{tabular}																					    
 \end{ruledtabular}																				    
 \end{table*}

\subsection{Decay scheme of $\bm{^{26}\mathrm{P}}$}

\begin{figure*}
\centering
 \includegraphics[trim=0mm 5mm 1cm 0mm, clip=true,width=\textwidth]{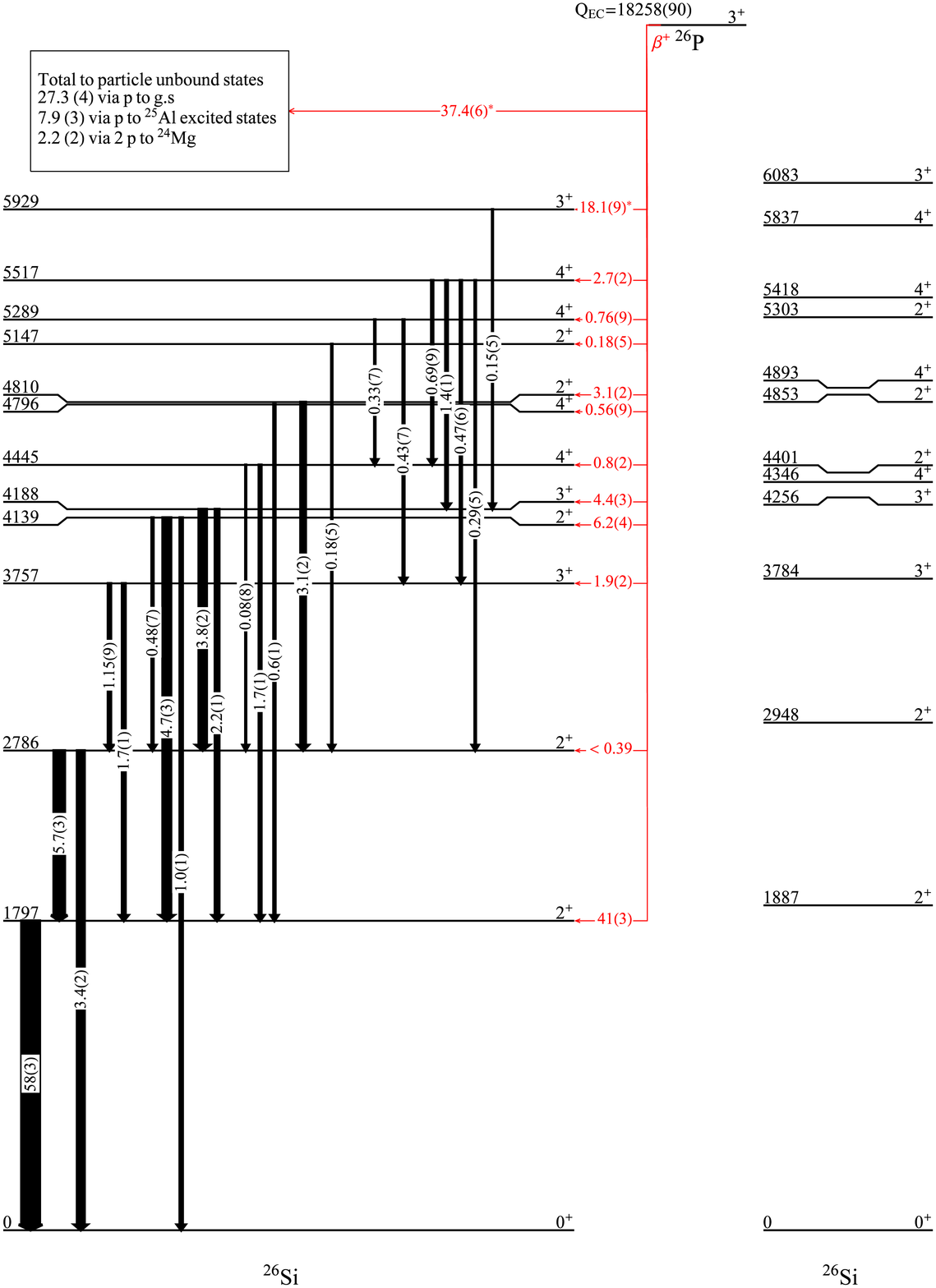}%
 \caption{\label{fig:decay} (Left) $^{26}$P decay scheme as deduced from the experimental data acquired in the present work. $\gamma$-ray transition labels correspond to the absolute intensities. $\beta$-decay branches corresponding to each populated level are also given (red). The branches to the unbound $3^+$ state and the particle unbound states (asterisks) were taken from literature \cite{Thomas2004,Schwartz2015}. (Right) $^{26}$Si levels populated in $^{26}$P $\beta$ decay obtained from a USDB shell-model calculation.  Level energies are given in keV.}
 \end{figure*}

Fig. \ref{fig:decay} displays the $^{26}\mathrm{P}$ $\beta$-decay scheme deduced from the results obtained in this experiment. Only those levels populated in the $\beta$ decay are represented. This level scheme was  built in a self-consistent way by taking into account the  $\gamma$-ray energies and intensities observed in the singles spectrum of Fig. \ref{fig:spec} and the $\beta$-$\gamma$-$\gamma$ coincidence spectra described in Sec. \ref{subsec:coincidences}.

The excitation energies of $^{26}\mathrm{Si}$  bound levels, their $\beta$-feedings, the energies of the $\gamma$ rays, and the absolute intensities measured in this work are shown in Table \ref{tab:levels}.

\subsubsection{$^{26}\mathrm{Si}$ level energies, spins and parities }

 Level energies of $^{26}\mathrm{Si}$ populated  in the $\beta$ delayed $\gamma$ decay of  $^{26}\mathrm{P}$  were obtained from the measured $\gamma$-ray energies  including a correction for the nuclear recoil. The excitation energy values of the  levels listed in Table \ref{tab:levels}  were obtained from the weighted average of all the possible $\gamma$-ray cascades coming from that level. To assign spins and parities we compared the deduced level scheme with USDB shell-model calculations and took into account $\beta$-decay angular momentum  selection rules, showing a 1 to 1 correspondence for all the levels  populated by allowed transitions, with a fair agreement in the level energies within theoretical uncertainties of a few hundred  keV (see Fig. \ref{fig:decay} ).

\subsubsection{ $\beta$-feedings}
 The $\beta$ branching ratio to the $i$-th excited energy level can be determined from the $\gamma$-ray intensities:

\begin{equation}
\label{eq:BR}
BR_i = I_{i,\text{out}}-I_{i,\text{in}},
\end{equation}
where $I_{i,\text{out}}(I_{i,\text{in}})$ represents the total $\gamma$-ray intensity  observed decaying out of (into) the $i$-th level. The $\beta$-decay branches deduced from this experiment are given in Table \ref{tab:BR}, where they are also compared to previous measurements of $^{26}\mathrm{P}$  $\beta$ decay \cite{Thomas2004}. To investigate  the possible  missing intensity from the Pandemonium effect \cite{Hardy1977}, we have used a shell-model calculation to estimate the  $\gamma$-ray intensities of all possible transitions from  bound states  feeding each particular level, and found them to be on the order of the uncertainty or (usually) much lower.
 \begin{table*}[t]
 \caption{\label{tab:BR} Comparison of the $\beta$-branches and $\log\! ft$ values obtained in the present work with previous determinations and shell-model calculations. Values previously unknown are indicated by a dash.}

 \begin{ruledtabular}

\begin{tabular}[c]{l d d d d d d}
\multicolumn{1}{c}{$E_x$ (keV)} & \multicolumn{3}{c}{ $\beta$-Branch (\%) }    &\multicolumn{3}{c}{ $\log\! ft$}\\
\cline{2-4}\cline{5-7}
& \multicolumn{1}{c}{Present work}&\multicolumn{1}{c}{Ref. \cite{Thomas2004} }   &\multicolumn{1}{c}{Theory}  &
\multicolumn{1}{c}{Present work}&\multicolumn{1}{c}{Ref. \cite{Thomas2004} }  &\multicolumn{1}{c}{Theory} \\
\hline
1797   & 41(3) 		  	      & 44(12)  & 47.22			&  4.89(3) & 4.89(17)			   &4.81  	\\       						  
2786   &\textless0.39   	      & 3.3(20) & 0.37			& \multicolumn{1}{c}{ $>$6.76} & 5.87(72)  &6.77  	\\
3757   & 1.9(2) 		      & 2.68(68)& 1.17 			& 5.94(4) & 5.81(15) 			   & 6.135	\\
3842   & \multicolumn{1}{c}{not obs.} & 1.68(47)& \multicolumn{1}{c}{--}& \multicolumn{1}{c}{not obs.} & 6.00(17) & \multicolumn{1}{c}{--}	 \\
4139   & 6.2(4) 		      & 1.78(75)& 2.97 			& 5.37(3) & 5.93(32) 			   & 5.634	\\
4188   & 4.4(3) 		      & 2.91(71)& 8.88 			& 5.51(3) & 5.71(14) 			   & 5.182	\\
4445   & 0.8(2) 		      &\multicolumn{1}{c}{--}& 1.11	& 6.23(8) &\multicolumn{1}{c}{--}  	   & 6.071	\\
4796   & 0.56(9) 		      &\multicolumn{1}{c}{--}&	0.06	&  6.31(7) &\multicolumn{1}{c}{--}	   & 7.274	\\
4810   & 3.1(2) 		      &\multicolumn{1}{c}{--}&	4.45	&  5.57(3) & \multicolumn{1}{c}{--}	   & 5.934	\\
5147   & 0.18(5) 		      &\multicolumn{1}{c}{--}& 0.03	&  6.7(1) &\multicolumn{1}{c}{--} 	   &7.474 	\\
5289   & 0.76(7) 		      &\multicolumn{1}{c}{--}&	0.60	& 6.09(6) & \multicolumn{1}{c}{--}	   &6.158 	\\
5517   & 2.7(2) 		      &\multicolumn{1}{c}{--}&	3.96	&  5.51(4) & \multicolumn{1}{c}{--}	   &5.262 	\\
5929   & 0.15(5)\footnote[1]{Only the $\gamma$ branch was measured} & 17.96(90)\footnote[2]{Only the proton branch was measured}&10.08 &  6.7(1)\footnotemark[1] & 4.60(3)\footnotemark[2] 	   &4.810      \\
 \end{tabular}
 \end{ruledtabular}
 \end{table*}

\section{ Discussion}

\subsection{Comparison to  previous values of  $\bm{^{26}\mathrm{Si}}$ level energies}

 We compare in Table  \ref{tab:energies} the  energies and the spins and parities deduced in this work with previous values available in the literature \cite{Thomas2004,PhysRevC.75.062801,Komatsubara2014,Doherty2015}. The results of Ref. \cite{Thomas2004} correspond to $\beta$ decay, thus the same levels are expected to be populated. We observed six  levels of  $^{26}\mathrm{Si}$ for the first time in the $\beta$ decay of  $^{26}\mathrm{P}$. These six levels were previously reported using  nuclear reactions  to populate them \cite{PhysRevC.75.062801,Komatsubara2014,Doherty2015}. The previously reported energies for these levels are in  good agreement with  the results obtained in this work. However, it is worth  mentioning a significant discrepancy (up to 6~keV) with energies obtained in Refs. \cite{PhysRevC.75.062801,Doherty2015} for the  two $\gamma$ rays  emitted from the $4_4^+$ state to the  $3_1^+$ and  $2_2^+$ states (1759.7 and 2729.9~keV, respectively). Despite these discrepancies in the $\gamma$-ray energies, the excitation energy of the level reported is in excellent agreement with our results. However, it should be noted that the $\gamma$-ray branching ratios are inconsistent for the 1759.7-keV transition.

 The 3842-keV level reported in \cite{Thomas2004} was not observed in the present work. In agreement with \cite{PhysRevC.75.062801,Komatsubara2014,Doherty2015} we show that  this level does not exist, as the 2045-keV $\gamma$ ray emitted from this level to the first excited state is not seen either  in the spectrum of Fig. \ref{fig:spec} nor the coincidence spectrum with the 1797-keV peak (Fig.~\ref{fig:Coincidences2}).

The 4810-keV level was  previously tentatively assigned to be a $2^+$ state, but this assignment was not clear, because of the proximity to another level at 4830~keV assigned as a $0^+$. The fact that the 2024-keV line  appears in the spectrum confirms that the spin and parity is $2^+,3^+$ or $4^+$. If this level was  $0^+$, the $\beta$-decay transition which populates this level would be second forbidden ($\Delta J=3$,$\Delta\pi=0$) and  highly suppressed.

 We  observed also the  two levels located  just above the proton separation energy ($S_p=5513.8$~keV). The first one corresponds to a $4^+$ state with an energy of 5517~keV. This level was also reported in Refs. \cite{PhysRevC.75.062801,Komatsubara2014}. The second level  at 5929~keV was previously observed in $\beta$-delayed proton emission  by Thomas \emph{et al.} \cite{Thomas2004} and more recently reported in our previous paper describing the present experiment \cite{Bennett2013}. The results presented here with the same set of data, but with an independent analysis, confirm the evidence for the observation of a $\gamma$ ray emitted from that level in the present experiment.

\begin{table*}
 \caption{\label{tab:energies} Excitation energies, spins and parities  of $^{26}\mathrm{Si}$ levels from the present work compared to previous $\gamma$-ray work and to the allowed $^{26}\mathrm{P}(\beta\gamma)$ transitions predicted by the shell model. Only the states below 6~MeV are listed. Values not observed  are indicated by a dash.}

 \begin{ruledtabular}

\begin{tabular}{c d c d c d c d c d c c}
\multicolumn{2}{c}{Present work  } &\multicolumn{2}{c}{ Ref. \cite{Thomas2004}} &\multicolumn{2}{c}{  Ref. \cite{PhysRevC.75.062801}} &  \multicolumn{2}{c}{  Ref. \cite{Komatsubara2014}}&  \multicolumn{2}{c}{  Ref. \cite{Doherty2015}}&\multicolumn{2}{c}{Theory}\\
\multicolumn{2}{c}{ $^{26}\mathrm{P}(\beta\gamma)$}&\multicolumn{2}{c}{ $^{26}\mathrm{P}(\beta\gamma)$}&\multicolumn{2}{c}{ $^{16}\mathrm{O}(^{12}\mathrm{C},2\mathrm{n}\gamma)$}&
\multicolumn{2}{c}{ $^{\mathrm{nat}}\mathrm{Mg}(^{3}\mathrm{He},\mathrm{n}\gamma)$}&\multicolumn{2}{c}{ $^{\mathrm{24}}\mathrm{Mg}(^{3}\mathrm{He},\mathrm{n}\gamma)$}&\multicolumn{2}{c}{ $^{26}\mathrm{P}(\beta\gamma)$}\\
\cline{1-2}\cline{3-4}\cline{5-6}\cline{7-8}\cline{9-10}\cline{11-12}
$J_n^\pi$ &\multicolumn{1}{c}{$E_x$ (keV)} &$J_n^\pi$ &\multicolumn{1}{c}{$E_x$ (keV)} &$J_n^\pi$ &\multicolumn{1}{c}{$E_x$ (keV)} & $J_n^\pi$ &\multicolumn{1}{c}{$E_x$ (keV)}& $J_n^\pi$ &\multicolumn{1}{c}{$E_x$ (keV)}& $J_n^\pi$ &\multicolumn{1}{c}{$E_x$ (keV)} \\  
\cline{1-2}\cline{3-4}\cline{5-6}\cline{7-8}\cline{9-10}\cline{11-12}
$2_1^+$ & 1797.1(3) & $2_1^+$ & 1795.9(2) & $2_1^+$ & 1797.3(1) & $2_1^+$ & 1797.4(4)&$2_1^+$ &1797.3(1) &$2_1^+$ &1887	     \\  
$2_2^+$ & 2786.4(3) & $2_2^+$ & 2783.5(4) & $2_2^+$ & 2786.4(2) & $2_2^+$ & 2786.8(6)&$2_2^+$ &2786.4(2) &$2_2^+$ &2948	     \\
\multicolumn{1}{c}{--}&\multicolumn{1}{c}{--}& \multicolumn{1}{c}{--}&\multicolumn{1}{c}{--}& $0_2^+$ & 3336.4(6) & $0_2^+$ & 3335.3(4)&$0_2^+$ & 3336.4(2)&\multicolumn{1}{c}{--}&\multicolumn{1}{c}{--}\\
$3_1^+$ & 3756.8(3) & $(3_1^+)$ & 3756(2) & $3_1^+$ & 3756.9(2) & $3_1^+$ & 3756.9(4)& $3_1^+$ & 3757.1(3)&$3_1^+$ &3784\\
\multicolumn{1}{c}{--}&\multicolumn{1}{c}{--}& $(4_1^+)$ & 3842(2) &\multicolumn{1}{c}{--}&\multicolumn{1}{c}{--}&\multicolumn{1}{c}{--}&\multicolumn{1}{c}{--}&\multicolumn{1}{c}{--}&\multicolumn{1}{c}{--}&\multicolumn{1}{c}{--}&\multicolumn{1}{c}{--}\\ 
$2_3^+$ & 4138.6(4) & $2_3^+$ & 4138(1) & $2_3^+$ & 4139.3(7) & $2_3^+$ & 4138.6(4)& $2_3^+$ & 4138.8(13)    &$2_3^+$ &4401\\ 
$3_2^+$ & 4187.6(4) & $3_2^+$ & 4184(1) & $3_2^+$ & 4187.1(3) & $3_2^+$ & 4187.4(4) & $3_2^+$ & 4187.2(4) &$3_2^+$ &4256\\
$4_1^+$ & 4445.1(4) & \multicolumn{1}{c}{--}&\multicolumn{1}{c}{--} & $4_1^+$ & 4446.2(4) & $4_1^+$ & 4445.2(4)& $4_1^+$ & 4445.5(12)   &$4_1^+$ &4346\\
$4_2^+$ & 4796.4(5) & \multicolumn{1}{c}{--}&\multicolumn{1}{c}{--} & $4_2^+$ & 4798.5(5) & $4_2^+$ & 4795.6(4) & $4_2^+$ & 4796.7(4)  &$4_2^+$ &4893\\
$2_4^+$ & 4810.4(4) & \multicolumn{1}{c}{--}&\multicolumn{1}{c}{--} & $(2_4^+)$ & 4810.7(6) &$(2_4^+)$ & 4808.8(4)&$2_4^+$ & 4811.9(4)&$2_4^+$ &4853\\  
\multicolumn{1}{c}{--}&\multicolumn{1}{c}{--}& \multicolumn{1}{c}{--}&\multicolumn{1}{c}{--}& $(0_3^+)$ & 4831.4(10) & $(0_3^+)$ & 4830.5(7) & $0_3^+$ & 4832.1(4)&\multicolumn{1}{c}{--}& \multicolumn{1}{c}{--}\\ 
$2_5^+$ & 5146.5(6) & \multicolumn{1}{c}{--}& \multicolumn{1}{c}{--}& $2_5^+$ & 5146.7(9) & $2_5^+$ & 5144.5(4) & $2_5^+$ & 5147.4(8)&$2_5^+$ &5303 \\
$4_3^+$ & 5288.9(4) & \multicolumn{1}{c}{--}& \multicolumn{1}{c}{--}& $4_3^+$ & 5288.2(5) & $4_3^+$ & 5285.4(7)& $4_3^+$ & 5288.5(7) &$4_3^+$ &5418 \\ 
$4_4^+$ & 5517.3(3) & \multicolumn{1}{c}{--}& \multicolumn{1}{c}{--}& $4_4^+$ & 5517.2(5) & $4_4^+$ & 5517.8(11)& $4_4^+$ & 5517.0(5)&$4_4^+$ &5837 \\
\multicolumn{1}{c}{--}&\multicolumn{1}{c}{--}& \multicolumn{1}{c}{--}&\multicolumn{1}{c}{--}& $1_1^+$ & 5677.0(17)& $1_1^+$ & 5673.6(10) & $1_1^+$ & 5675.9(11)&\multicolumn{1}{c}{--}& \multicolumn{1}{c}{--} \\
\multicolumn{1}{c}{--}&\multicolumn{1}{c}{--}& \multicolumn{1}{c}{--}&\multicolumn{1}{c}{--}&\multicolumn{1}{c}{--}&\multicolumn{1}{c}{--} & $0_4^+$ & 5890.0(10)& $0_4^+$ & 5890.1(6)& \multicolumn{1}{c}{--}& \multicolumn{1}{c}{--}\\
$3_3^+$ & 5929.3(6) & $3_1^+$ & 5929(5)\footnote{$^{26}\mathrm{P}(\beta\mathrm{p})$.} & \multicolumn{1}{c}{--}&\multicolumn{1}{c}{--}&\multicolumn{1}{c}{--}&\multicolumn{1}{c}{--}&\multicolumn{1}{c}{--}&\multicolumn{1}{c}{--}&$3_3^+$ &6083 \\ 

\end{tabular} 
\end{ruledtabular}
\end{table*}

\subsection{$\bm{ft}$ values and Gamow-Teller strength}
As mentioned in Sec. \ref{sec:intro}, the calculation of the experimental $ft$ values requires the measurement of three fundamental quantities: (a) the half-life, (b) the branching-ratio, and (c) the $Q$ value of the decay. The experimental value of the half-life and  the semiempirical $Q$-value, are $t_{1/2}=43.7(6)$~ms and $Q_{EC}=18250(90)$~keV, respectively. Both values were taken from Ref. \cite{Thomas2004}. The branching ratios from the present work  are listed in Table \ref{tab:levels}. The partial half-lives $t_i$, are thus calculated as:
\begin{equation}
\label{eq:partial_half_life}
t_i = \frac{t_{1/2}}{BR_i}(1+P_{EC}),
\end{equation}

 where $BR_i$ is the $\beta$-branching ratio of  the $i$-th level and $P_{EC}$ the fraction of electron capture, which can be neglected for the light nuclide $^{26}$P. The statistical  phase space factors $f$ were calculated with the parametrization reported in \cite{Wilkinson197458} including  additional radiative  \cite{WILKINSON1973_2} and diffuseness corrections \cite{PhysRevC.18.401}. The uncertainty associated with this calculation is 0.1\%, which is added quadratically to the uncertainty derived from the 0.5\% uncertainty of the $Q_{EC}$ value. Table \ref{tab:BR} shows the $\beta$ branches and $\log\!ft$ values for the transitions to excited levels of $^{26}$Si  compared to the previous values reported in \cite{Thomas2004}. For the first excited  state, our estimation of the $\beta$ feeding  is consistent with the previous result. In the case of the second excited state, the previous value is one order of magnitude larger than our upper limit. This  is because of the new levels we  observed. The large branching ratios observed for the $2_3^+$ and the $3_2^+$ states compared to previous results, 6.2(4)\% and 4.4(3)\%, respectively, are noteworthy. The reason for that difference is the observation of new $\gamma$ rays emitted by those levels which have now been accounted for. The new levels together with the unobserved state at 3842~keV explain all the discrepancies between the results reported here and literature values \cite{Thomas2004}.
 As far as the $\log\! ft$ values are concerned the agreement for the first excited state is very good, but when going to higher energies, the discrepancies in the $\log\! ft$ values are directly related to those in the branching ratios.

\subsubsection{Comparison to theory}

Theoretical calculations were also performed using a shell model code. Wave functions of $^{26}$P were deduced using a full $sd$-shell model with the USDB interaction and their corresponding beta decay transitions to $^{26}$Si levels.

Fig. \ref{fig:decay} shows the comparison  between the $^{26}\mathrm{Si}$ level energies  deduced in this $^{26}$P $\beta$-decay work to the same levels predicted by the calculation. We observe a fair  agreement in the level energies, but the theoretical values are systematically higher. The r.m.s. and maximum deviations between theory and experimental results are 109 and 320~keV, respectively. From a direct comparison we also see that in this work we have measured all the states populated in the allowed transitions predicted by the shell-model calculation.

The experimental $\log\! ft$ values presented in Table \ref{tab:BR} were determined from the measured  branching ratios combined with the known values of $Q_{EC}$ and half-life \cite{Thomas2004}.  Theoretical Gamow-Teller strengths were obtained from the matrix elements of the transitions to states of $^{26}$Si populated in the $\beta$ decay of $^{26}$P. To compare them to the experimental results, the experimental $B(GT)$ values were calculated from the   $ft$ values through the expression,

\begin{equation}  \label{eq:BGT}
 B(GT)=\frac{2\mathcal{F}t}{ft}, 
\end{equation}

\begin{table}[b]
  \caption{Comparison of the experimental and theoretical $B(GT)$ values obtained in the present work. The quenching factor applied to theory is $q^2=0.6$.}
  \label{tab:bgt}
 \begin{ruledtabular}
  \begin{tabular}{d  d c d d }
\multicolumn{2}{c}{Present work}& &\multicolumn{2}{c}{Theory}\\
\cline{1-2} \cline{4-5}   
\multicolumn{1}{c}{$E_x$ (keV)} & \multicolumn{1}{c}{ $B(GT)$}& $I_n^\pi$ &
\multicolumn{1}{c}{$E_x$ (keV)}   & \multicolumn{1}{c}{ $B(GT)$}
\\
\hline
1797  & 0.048(3)		&$2_1^+$ &1887   &0.0606 \\	
2786  &\textless0.0007  	&$2_2^+$ &2948   &0.0007 \\
3757  & 0.0044(4)		&$3_1^+$ &3784   & 0.0029 \\
4139  & 0.016(1)		&$2_3^+$ &4401   & 0.009 \\
4188  & 0.0117(1)		&$3_2^+$ &4256   & 0.0256\\
4445  & 0.0023(4)		&$4_1^+$ &4346   & 0.0033\\
4796  & 0.0018(3)		&$4_2^+$ &4893   & 0.0002\\
4810  & 0.0103(7)		&$2_4^+$ &4853   & 0.0161\\
5147  & 0.0007(2)		&$2_5^+$ &5303   &0.0001  \\
5289  & 0.0031(4)		&$4_3^+$ &5418   & 0.0027\\
5517  & 0.012(1)		&$4_4^+$ &5837   & 0.0213\\

  \end{tabular}
 \end{ruledtabular}
\end{table}

where $\mathcal{F}t= 3072.27\pm0.62$~s \cite{Hardy2015} is the average corrected $ft$ value from $T=1$ $0^+\rightarrow 0^+$ superallowed  Fermi $\beta$ decays. Table \ref{tab:bgt} shows the comparison between the  experimental and theoretical $B(GT)$ values. A quenching factor $q=0.77$ ($q^2=0.6$) was applied to the shell-model calculation \cite{Wildenthal1983}. Theoretical predictions overestimate the experimental values for  the transitions to the $2^+_1$, $3^+_2$, $4^+_1$, $2^+_4$, and $4^+_4$ states. Experimental $B(GT)$ values are slightly underestimated for the rest of the states up to 5.9~MeV.   The most significant differences are in the $4_2^+$  and the $2_5^+$ levels for which the  predicted $B(GT)$ values differ by almost  one order of magnitude with the experimental ones. A possible explanation for this difference is the mixing between different levels. 

\begin{figure}[t]
 \includegraphics[trim=0mm 5mm 2mm 0mm, clip=true,width=.45\textwidth]{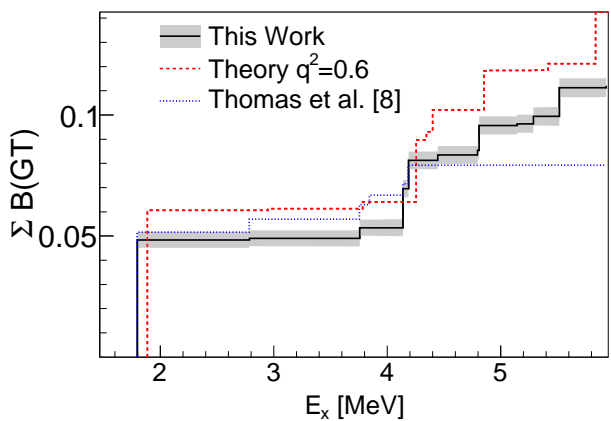}%
 \caption{\label{fig:BGT}  Summed Gamow-Teller strength distribution of the $\beta$ decay of $^{26}$P up to 5.9~MeV excitation energy. The results of the present experiment are compared to previous results \cite{Thomas2004} and Shell-Model calculations. A quenching factor $q^2=0.6$ was used in the theoretical calculation.}
 \end{figure}

\begin{table*}
  \caption{Comparison of experimental $ft$ values for the $\beta$ decay of $^{26}$P and its mirror $^{26}$Na \cite{PhysRevC.71.044309}. The mirror asymmetry $\delta$ is also listed and compared to the previous experimental results \cite{Thomas2004}, where applicable.}
  \label{tab:mirror}
 \begin{ruledtabular}
  \begin{tabular}{d c  c  d c d d}
\multicolumn{2}{c}{ $^{26}\mathrm{P}(\beta\gamma)^{26}\mathrm{Si}$}& &\multicolumn{2}{c}{ $^{26}\mathrm{Na}(\beta\gamma)^{26}\mathrm{Mg}$ \cite{PhysRevC.71.044309}}& \multicolumn{2}{c}{ $\delta (\%)$}\\
\cline{1-2} \cline{4-5} \cline{6-7}  
\multicolumn{1}{c}{$^{26}\mathrm{Si}\;E_x$ (keV)}  & \multicolumn{1}{c}{$ft^+$ (s) } & $I_n^\pi$& 
\multicolumn{1}{c}{$^{26}\mathrm{Mg}\;E_x$ (keV)}  & \multicolumn{1}{c}{$ft^-$ (s) } & \multicolumn{1}{c}{Present work}&\multicolumn{1}{c}{Ref.\cite{Thomas2004}}
\\
\hline
1797  & 7.9(5)$\times 10^4$ 	& $2_1^+$ &1809 &5.23(2)$\times 10^4$  & 51(10) &50(60)\\       
3757  & 8.7(8)$\times 10^5$ 	& $3_1^+$ &3941  &7.5(2)$\times 10^5$  &16(11)&10(40)  \\
4139  & 2.4(2)$\times 10^5$ 	& $2_3^+$ &4332  &4.22(9)$\times 10^5$  &-43(5)&110(160)  \\
4188  & 3.2(2)$\times 10^5$  	& $3_2^+$ &4350  &2.16(4)$\times 10^5$   & 50(10) &110(70)\\
4445  & 1.7(7)$\times 10^6$ 	& $4_1^+$ &4319  &1.43(3)$\times 10^6$  &20(50) \\
4796  & 2.1(3)$\times 10^6$ 	& $4_2^+$ &4901  &1.63(7)$\times 10^6$  & 29(18)\\
4810  & 3.7(3)$\times 10^5$ 	& $2_4^+$ &4835  &1.85(2)$\times 10^5$  & 100(16)\\
5147  & 5.6(20)$\times 10^6$  	& $2_5^+$ &5291  &2.0(3)$\times 10^7$ &  -72(11)\\
5289  & 1.2(2)$\times 10^6$ 	& $4_3^+$ &5476  &7.9(40)$\times 10^7$ & -98(1)\\
5517  & 3.2(3)$\times 10^5$  	& $4_4^+$ &5716&  1.71(3)$\times 10^5$&87(18) \\


  \end{tabular}
 \end{ruledtabular}
\end{table*}

 Fig. \ref{fig:BGT} shows the summed Gamow-Teller strength distribution of the decay of $^{26}$P for  bound levels up to 5517~keV. In this figure we compare the results obtained in this work with the previous results and the shell-model calculation.  We can see that the agreement with the previous experimental results is good for the first excited state, with a small difference that is consistent within uncertainties. As the energy increases the differences become more significant, with our results slightly below the previous ones until the contribution of the new levels is added. For energies above 4.1 MeV, the results from the previous experiment are clearly below our results. If we compare the present data with the theoretical prediction using the typical quenching factor of $q^2=0.6$, we see that the theoretical prediction overestimates the summed Gamow-Teller strength in the excitation energy region below 5.9~MeV. 
If a quenching factor of 0.47 were applied to the shell model calculations instead, the agreement would be almost perfect in this energy region. However, this does not necessarily imply that the value of $q^2=0.6$ is inapplicable because only a small energy range was considered for the normalization. In fact, most of  the Gamow-Teller strength is to unbound states which have not been measured in the present work. Furthermore, according to shell model calculations, only $\sim$21\% of the total Gamow-Teller strength is in the $Q$-value window.

\subsection{Mirror asymmetry and $\bm{^{26}\mathrm{P}}$ proton halo}

The high precision data on the $\beta$ decay of the mirror-nucleus $^{26}$Na from Ref. \cite{PhysRevC.71.044309}, together with the results obtained in the present work made it possible to calculate finite values of the mirror asymmetry for $\beta$-decay transitions from the  $A=26$, $T_z=\pm2$ mirror nuclei to low lying states of their respective daughters. Table \ref{tab:mirror} shows the results of the $ft$ values obtained for the $\beta$ decay of $^{26}$P and its mirror nucleus, and the corresponding asymmetry parameter,  compared to the previous experimental results reported in Ref. \cite{Thomas2004}. We see that for the low lying  states, the agreement between previous data and our results is good, but our results are  more precise, yielding the first finite values for this system. For the higher energy states, we report the first values for the mirror asymmetry. We observe  large and significant   mirror asymmetries with values ranging from $-98\%$ up to $+100\%$. As mentioned in Sec. \ref{sec:intro}, mirror asymmetries can be related to isospin mixing and/or differences in the radial wavefunctions. It was also shown that halo states produce  significant mirror asymmetries. The $51(10)\%$ asymmetry observed for the transition to the first excited state could be further evidence for a proton halo in $^{26}$P \cite{Navin1998}. Higher lying states are not as useful because of possible mixing between nearby states.

To investigate this effect more quantitatively, we performed  two different shell model calculations with the USDA and USDB interactions. For the transition to the first excited state, these two interactions  predict mirror asymmetries of 3\% and 2.5\%, respectively: far from  experimental result. If we lower the energy of the $2s_{1/2}$ proton orbital by 1~MeV  to account for the low proton separation energy of $^{26}$P, the mirror asymmetries we obtain  for the first excited state are 60\%  and 50\% for the USDA and USDB interactions, respectively, in agreement with the experimental result and supporting the  hypothesis of a halo state \cite{Brown1996}. Before firm conclusions can be made, however, more detailed calculations are needed to evaluate the contributions of the  other effects that may produce mirror asymmetries.

\section{ $\bm{^{25}\mathrm{Al}(\mathrm{p},\gamma)^{26}\mathrm{Si}}$ Reaction rate calculation}

As reported in Ref. \cite{Wrede_2009}, the $\beta$ decay of $^{26}$P to $^{26}$Si provides a convenient means for determining parameters of the astrophysically relevant reaction $^{25}$Al$(p,\gamma)^{26}$Si  in novae. In these stellar environments, the nuclei are assumed to have a Maxwell-Boltzmann distribution of energies characterized by the temperature $T$ from which the  resonant reaction rate can be described by a sum over the different resonances:

\begin{equation}
\label{eq:Reaction rate}
\langle \sigma v\rangle=\left (\frac{2\pi}{\mu kT}\right )^{3/2}\hbar^2\sum_r(\omega\gamma)_re^{-E_r/kT},
\end{equation}

where $\hbar$ is the reduced Planck constant, $k$ is the Boltzmann constant, $\mu$ is the reduced mass, and $E_r$ is the energy of the resonance in the center-of-mass frame. $(\omega\gamma)_r$ is the resonance strength, which is  defined as

\begin{equation}
\label{eq:Res_stength}
(\omega\gamma)_r=\frac{(2J_r+1)}{(2J_p+1)(2J_{\text{Al}}+1)}\left(\frac{\Gamma_p\Gamma_\gamma}{\Gamma} \right )_r.
\end{equation}

\begin{table} 

  \caption{Thermonuclear $^{25}$Al$(p,\gamma)^{26}$Si reaction rate, $N_A\langle\sigma v\rangle$, in units of cm$^3$s$^{-1}$mol$^{-1}$ as a function of stellar temperature $T$, including resonant capture contributions from  resonances and direct capture. The first and last columns labeled ``Low'' and ``High'', respectively, correspond to the 1 standard deviation uncertainty limits, while the ``Central'' one corresponds to the recommended rate.}
  \label{tab:rate}


 \begin{ruledtabular}
  \begin{tabular}{ d n{2}{2} n{2}{2} n{2}{2} }

\multicolumn{1}{c}{$T$ (GK)}	&	\multicolumn{1}{c}{Low}		&       	\multicolumn{1}{c}{Central}&       	\multicolumn{1}{c}{High}	     \\
\hline
0.01	&	1.10E-37	&	1.57E-37	&	2.04E-37	\\
0.015	&	7.00E-32	&	1.00E-31	&	1.30E-31	\\
0.02	&	3.19E-28	&	4.56E-28	&	5.93E-28	\\
0.03	&	1.23E-23	&	1.75E-23	&	2.28E-23	\\
0.04	&	9.42E-21	&	1.34E-20	&	1.75E-20	\\
0.05	&	1.40E-18	&	1.93E-18	&	2.88E-18	\\
0.06	&	1.16E-16	&	2.42E-16	&	6.17E-16	\\
0.07	&	5.64E-15	&	1.50E-14	&	4.30E-14	\\
0.08	&	1.27E-13	&	3.59E-13	&	1.06E-12	\\
0.09	&	1.46E-12	&	4.23E-12	&	1.25E-11	\\
0.1	&	1.03E-11	&	3.01E-11	&	8.95E-11	\\
0.11	&	5.06E-11	&	1.48E-10	&	4.40E-10	\\
0.12	&	1.99E-10	&	5.53E-10	&	1.64E-09	\\
0.13	&	5.80E-10	&	1.68E-09	&	4.98E-09	\\
0.14	&	1.55E-09	&	4.36E-09	&	1.28E-08	\\
0.15	&	4.04E-09	&	1.03E-08	&	2.92E-08	\\
0.16	&	1.14E-08	&	2.43E-08	&	6.24E-08	\\
0.17	&	3.46E-08	&	6.23E-08	&	1.34E-07	\\
0.18	&	1.02E-07	&	1.79E-07	&	3.14E-07	\\
0.19	&	2.84E-07	&	5.41E-07	&	8.44E-07	\\
0.2	&	7.80E-07	&	1.60E-06	&	2.42E-06	\\
0.21	&	2.07E-06	&	4.47E-06	&	6.75E-06	\\
0.22	&	5.21E-06	&	1.15E-05	&	1.75E-05	\\
0.23	&	1.23E-05	&	2.76E-05	&	4.21E-05	\\
0.24	&	2.72E-05	&	6.17E-05	&	9.40E-05	\\
0.25	&	5.67E-05	&	1.29E-04	&	1.97E-04	\\
0.26	&	1.12E-04	&	2.55E-04	&	3.89E-04	\\
0.27	&	2.09E-04	&	4.78E-04	&	7.30E-04	\\
0.28	&	3.74E-04	&	8.55E-04	&	1.31E-03	\\
0.29	&	6.42E-04	&	1.47E-03	&	2.24E-03	\\
0.3	&	1.06E-03	&	2.43E-03	&	3.71E-03	\\
0.31	&	1.70E-03	&	3.88E-03	&	5.93E-03	\\
0.32	&	2.63E-03	&	6.01E-03	&	9.19E-03	\\
0.33	&	3.96E-03	&	9.06E-03	&	1.39E-02	\\
0.34	&	5.82E-03	&	1.33E-02	&	2.04E-02	\\
0.35	&	8.36E-03	&	1.91E-02	&	2.92E-02	\\
0.36	&	1.18E-02	&	2.69E-02	&	4.10E-02	\\
0.37	&	1.62E-02	&	3.70E-02	&	5.66E-02	\\
0.38	&	2.19E-02	&	5.01E-02	&	7.66E-02	\\
0.39	&	2.92E-02	&	6.67E-02	&	1.02E-01	\\
0.4	&	3.83E-02	&	8.75E-02	&	1.34E-01	\\
0.42	&	6.32E-02	&	1.44E-01	&	2.21E-01	\\
0.44	&	9.94E-02	&	2.27E-01	&	3.47E-01	\\
0.46	&	1.50E-01	&	3.42E-01	&	5.22E-01	\\
0.48	&	2.17E-01	&	4.96E-01	&	7.58E-01	\\
0.5	&	3.06E-01	&	6.97E-01	&	1.06E+00	\\

  \end{tabular}
 \end{ruledtabular}
\end{table}

$J_{r(p,\mathrm{Al})}$ are the spins of the resonance (reactants), $\Gamma_{p(\gamma)}$ are the proton ($\gamma$-ray) partial widths of the resonance and $\Gamma=\Gamma_p+\Gamma_\gamma$ is the total width. It was previously predicted \cite{Iliadis_96} that the levels corresponding to significant resonances  at nova temperatures in  the $^{25}$Al$(p,\gamma)^{26}$Si reaction are the $J^\pi = 1_1^+,4_4^+,0_4^+$, and $3_3^+$  levels. In our previous work \cite{Bennett2013} we reported the first evidence for the observation of $\gamma$ rays emitted from the $3_3^+$ level. The determination of the strength of the $3_3^+$ resonance in $^{25}$Al$(p,\gamma)^{26}$Si based on the experimental measurements of the partial proton width  ($\Gamma_p$) \cite{Peplowski2009} and the  $\gamma$-ray branching ratio ($\Gamma_\gamma/\Gamma$) \cite{Bennett2013}  was also performed and used  to determine the amount of $^{26}$Al ejected in novae. In this work, we have confirmed the evidence for the 1742-keV $\gamma$ ray emitted from the  $3_3^+$ level to the $3_2^+$ level in $^{26}$Si with an intensity of $0.15(5)\%$. To some extent, the present paper is a follow-up of our previous work, thus we present here (see Table \ref{tab:rate}) for completeness the results of the full reaction rate calculation used to obtain the astrophysical results published in \cite{Bennett2013}. The table shows the total thermonuclear $^{25}$Al$(p,\gamma)^{26}$Si reaction rate as a function of temperature including contributions from the relevant resonances, namely $1_1^+,0_4^+$, and $3_3^+$  and the direct capture. For the  $1^+$ and $0^+$ resonances and the direct capture, values are adopted from Ref. \cite{Wrede_2009}. Our table includes the rate limits calculated from a 1 standard deviation variation of the parameters.

\section{Conclusions}

We have measured the absolute $\gamma$-ray intensities and deduced the $\beta$-decay branches for the decay of $^{26}$P to bound states and low-lying  resonances of $^{26}$Si. We have observed six new $\beta$-decay branches and 15 $\gamma$-ray lines never observed before in $^{26}$P  $\beta$ decay, likely corresponding to most of all the allowed  Gamow-Teller transitions between the ground state and 5.9~MeV. The energies measured for the excited states show good agreement with previous results obtained using various nuclear reactions to populate these states. We have calculated the $\log\! ft$ values of all these new transitions and compared them to USDB shell-model calculations. The reported values  show  good agreement with the theoretical calculations. In addition, the Gamow-Teller strength function was calculated and compared to theoretical values, showing that the  summed Gamow Teller strength is locally overestimated  with the standard  $sd$ shell quenching of 0.6. The mirror asymmetry was also investigated by calculating the $\beta$-decay asymmetry parameter $\delta$ for 10 transitions. The significant asymmetries observed, particularly for the transition to the first excited states of $^{26}$Si and its mirror $^{26}$Mg ($\delta=(51\pm10)\%$) might be further evidence for the existence of a proton halo in the $^{26}$P.  Finally, we have tabulated the total $^{25}$Al$(p,\gamma)^{26}$Si reaction rate at nova temperatures used to estimate the galactic production of $^{26}$Al in  novae in Ref. \cite{Bennett2013}.

\begin{acknowledgments}
The authors gratefully acknowledge the contributions of the NSCL staff.
This work is supported by the U.S. National Science Foundation under grants PHY-1102511,  PHY-0822648, PHY-1350234, PHY-1404442, the U.S. Department of Energy under contract No. DE-FG02-97ER41020, the U.S. National Nuclear Security Agency under contract No. DE-NA0000979 and the Natural Sciences and Engineering Research Council of Canada.
\end{acknowledgments}

\providecommand{\noopsort}[1]{}\providecommand{\singleletter}[1]{#1}%

\end{document}